\documentclass[12pt,preprint]{aastex}
\usepackage[dvips]{color}

\slugcomment{draft version \today, }

\shorttitle{Pop III GRBs and
Breakout Criteria for Accretion-Powered Jets}
\shortauthors{Nagakura et al.}

\begin{document}

\title{
Population III Gamma-Ray Bursts
and Breakout Criteria for Accretion-Powered Jets
}
%Resolving the $z=4.4$ Quasar Host Galaxy of BRI\,1335-0417: \\
%Intimate Interaction or Massive Molecular Disk?}

\author{Hiroki Nagakura\altaffilmark{1,2}, Yudai Suwa\altaffilmark{1}, Kunihito Ioka\altaffilmark{3,4}} 

\altaffiltext{1}{Yukawa Institute for Theoretical Physics, Kyoto
  University, Oiwake-cho, Kitashirakawa, Sakyo-ku, Kyoto, 606-8502,
  Japan}
\altaffiltext{2}{Department of Science and Engineering, Waseda
  University, 3-4-1 Okubo, Shinjuku, Tokyo 169-8555, Japan}
\email{hiroki@heap.phys.waseda.ac.jp}
\altaffiltext{3}{KEK Theory Center, 1-1 Oho, Tsukuba 305-0801, Japan}
\altaffiltext{4}{Department of Particle and Nuclear Physics, The Graduate University for Advanced Studies (Sokendai), 1-1 Oho, Tsukuba 305-0801, Japan}

%\altaffiltext{3}{KEK Theory Center and the Graduate University for
%  Advanced Studies (Sokendai), 1-1 Oho, Tsukuba 305-0801, Japan}

\begin{abstract}
We investigate the propagation of accretion-powered jets in various
types of massive stars such as Wolf-Rayet stars, light Population III
(Pop III) stars, and massive Pop III stars, all of which are the
progenitor candidates of Gamma-Ray Bursts (GRBs). 
We perform two dimensional axisymmetric simulations of
relativistic hydrodynamics taking into account both the envelope
collapse and the jet propagation (i.e., the negative feedback of the
jet on the accretion).
 Based on our hydrodynamic simulations, we show for the first time that
 the accretion-powered jet can potentially break out relativistically
 from the outer layers of Pop III progenitors.
 In our simulations, the accretion rate is estimated by the mass flux
 going through the inner boundary,
 and the jet is injected with a fixed accretion-to-jet conversion efficiency $\eta$.
%We calculate the accretion rate from the mass flux going through the
%inner boundary,
% and inject the jet with a constant accretion-to-jet
%conversion efficiency $\eta$.
By varying the efficiency $\eta$ and
opening angle $\theta_{op}$
%\naga{for $\sim 40$ models}, 
for more than $40$ models,
we find that the jet can make a relativistic breakout from all types of progenitors for GRBs if a simple condition $\eta \gtrsim 10^{-4}
(\theta_{op}/8^{\circ})^2$ is satisfied, which is consistent with
analytical estimates. Otherwise no explosion or some failed spherical
explosions occur.
%We also suggest that a (slightly) late time operation of the central engine is preferred for GRBs.
%\naga{We suggest that the high efficient accretion-powered jet can potentially create GRBs, no matter if progenitors of GRBs have the massive stellar envelope.}
%the GRB jets.
\end{abstract}

\keywords{black hole physics, hydrodynamics}

\section{Introduction}
The link between nearby Gamma Ray Bursts (GRBs) and
peculiar Type Ib/c supernovae (or hypernovae) ambiguously shows that
some populations of GRBs are born from the catastrophic death of
massive stars. Observations of host galaxies of GRBs also lead to the
general consensus that GRBs are generated preferentially in low
metallicity star-forming regions \citep[see
  e.g.][]{2008AJ....135.1136M}. The stellar evolutions at low
metallicity are believed to suppress the mass loss, so that the star
can maintain its own angular momentum. As a result, the iron core of
these stars would be rapidly spinning \citep[see
  e.g.][]{2005A&A...443..643Y,2006ApJ...637..914W}, which would be a
necessary condition for producing GRBs in the collapsar
\citep{1993ApJ...405..273W,1999ApJ...524..262M} or magnetar models
\citep[see e.g.][]{2004ApJ...611..380T,2011MNRAS.413.2031M}. According
to these facts, it is widely recognized that rapidly rotating
Wolf-Rayet stars in low metallicity regions are the most favored
progenitors for GRBs.

The first stars (hereafter Pop III) also potentially create GRBs. The first stars are supposed to be formed with a huge mass ($M \gtrsim 100 M_{\sun}$)
\citep{2002Sci...295...93A,2002ApJ...564...23B} and a rapid rotation
with nearly breakup speed \citep{2011MNRAS.tmp..142S}. The
gravitational collapse of these stars would result in a black hole
formation
\citep{2006ApJ...645..519N,2007PASJ...59..771S,2009ApJ...690..913S,2011ApJ...737....6S}
and potentially the central engine activity
\citep{2003ApJ...591..288H,2010ApJ...715..967M,2010MNRAS.402L..25K,2011ApJ...726..107S}.
If the primordial gas is ionized by radiation from first-generation metal-free (Pop III.1) stars, the subsequent metal-free (Pop III.2) stars would be less massive \citep{2009Natur.459...49B} and outnumber the Pop III.1 stars \citep{2011A&A...533A..32D}.
It has also been recently discussed that the radiative feedback could reduce the mass of the first star via the
HII region breakout and the photoevaporation of the accretion disk
\citep{2008ApJ...681..771M,2011Sci...334.1250H}.
%(McKee and Tan 2008, Hosokawa and Omukai private 2010 ???  please add
%references).  Recently, {\bf Hosokawa et al.} found that the mass
%accretion onto proto Pop III star was supplesed by radiation
%feedback, then the outcomes of mass at ZAMS became $M \sim 45
%M_{\sun}$.
Although the reduced mass $\lesssim 100 M_{\odot}$ is significantly lower than
previously thought, the light Pop III stars still have large enough 
mantles for the formation of a black hole. Thus, even in these cases,
the central engine could operate as a result of the core collapse in
the standard collapsar model.  The Pop III GRBs and their afterglows
are detectable in principle up to $z \sim 100$ and $z \sim 30$,
respectively, providing powerful probes of the high redshift universe
  \citep{2000ApJ...536....1L,2000ApJ...540..687C,2003ApJ...598L..79I,2004ApJ...604..508G,2005ApJ...619..684I,
    2007MNRAS.380.1715I,2011ApJ...731..127T}.
%(Ioka and Meszaros 2005, Inoue et al. 2007, Toma et al. 2010
%??? please add references ???).

However, even if the central engine successfully operates, it is still
a matter of debate whether the jet can produce GRBs or not. One of the
main obstacles for producing GRBs is the stellar envelope that may
prevent the jet propagation. If the central engine turns off well
before the jet head reaches the stellar surface, all of the jet matter
undergoes dissipation by the reverse shock wave and it will
eventually expand spherically. In addition, the outflow is
contaminated by a huge amount of baryons, so it is naturally expected
that its velocity becomes non-relativistic and never create a GRB. 
With this expectation, \citet{2003MNRAS.345..575M} constrained the
progenitors of GRBs assuming that the lifetime of the central engine
is comparable to the observed duration of the prompt phase of
GRBs. He concluded that only compact carbon-oxygen Wolf-Rayet stars
satisfy the condition for producing GRBs, while very massive
stars such as Pop III stars are not suitable.

On the contrary, \citet{2011ApJ...726..107S} recently pointed out that
the jet breakout is possible even if the Pop III star has a supergiant
hydrogen envelope without mass loss, thanks to the long-lived powerful
accretion of the envelope itself.  They analytically showed that the jet
successfully penetrates the Pop III as well as compact Wolf-Rayet
stars if the envelope continues to fall in a black hole and the
accretion-to-jet conversion efficiency is larger than a certain level.

However, it is not trivial to determine whether the envelope can continue to fall in and accrete onto black holes or not. Generally, the core collapse produces
rarefaction waves, which propagate outwards through the envelope and
induce the infall of the stellar envelope
\citep[see][]{2011ApJ...731...80N}. 
%It should be noted that 
But some portions of the envelope cease to fall due to the jet propagation
when the central engine begins to operate.
%, but little is known about how these
%jet dynamics give the negative feedback on the accretion to a black hole.  
Although almost all matter could accrete from the
equatorial regions, the feedback would affect the accretion
rate if the jet opening angle is large. 
The jet feedback to the accretion has not been taken into
account
in previous studies
\citep{2003MNRAS.345..575M,2008ApJ...675..519J,2008MNRAS.388.1729K,2010ApJ...713..800L,2011ApJ...726..107S}.
Since these processes are supposed to be complex and strongly
non-linear phenomena, hydrodynamic simulations with both
accretions and jet propgations are strongly required.

% It is because very light jets, on which GRB jet is based, create the
%strong forward shock and also hot cocoon, so that they make some
%portions of matter go outwards, then it may cease the accretions.
%{\bf Since the light jet creates the forward shock and hot cocoon,
%infalling matter induced by the rarefaction wave can not directly
%accrete onto a black hole.}

On the other hand, a large number of numerical studies on jet
propagations in the stellar mantle have been carried out
\citep{1999ApJ...524..262M,2006ApJ...651..960M,2007ApJ...665..569M,2007ApJ...657L..77T,2009ApJ...700L..47L,2009ApJ...699.1261M,2011ApJ...732...26M,2011ApJ...731...80N}.
However, almost all works assume that the jet is injected with a
constant energy flux from a certain radius of the inner
boundary. Although we do not know the mechanism of the central
engine, it is naturally expected that the jet luminosity would
correlate with the accretion rate in one way or another
\citep{2002ApJ...579..706D,2003ApJ...599L...5P,2006MNRAS.368.1561M,2011MNRAS.410.2302Z} (but see also \citep{2011MNRAS.413.2031M} for a magnetar model that does not depend on accretion).
We also wonder whether the jet production could be ceased by the reduction of the
accretion due to the negative feedback as mentioned above.
In the previous study, \citet{2001ApJ...550..410M} demonstrated the jet
propagation with the jet luminosity as a function of the accretion
rates. However, they calculated the jet propagation and the fall back
process separately, and can not address the jet feedback process
adequately. In another previous study, \citet{2010ApJ...723..267M}
investigated time variable jet injections, but the
luminosity and time variability are determined by hand. In order to
investigate the jet propagation feedback on the
accretion process, it is necessary to perform numerical
simulations in both the collapsing phase and the jet propagation phase
at once.

% Note also that they initially induced spherical explosions, then the
% fall back accretion rate was estimated (see more details in their
% original paper).

Motivated by these facts, we perform two-dimensional axisymmetric
hydrodynamic simulations for the envelope collapse and the jet
propagation in a single computation. The purpose of this study is to
clarify whether the forward shock wave successfully propagates
and breaks out from the various types of the stellar progenitors for
GRBs or not, taking into account the jet feedback process. 
We survey the parameter space of the jet opening angle
and the accretion-to-jet conversion efficiency,
and discuss how these key quantities affect the jet dynamics to obtain
the simple analytical criteria for the GRB production.
%\naga{We also perform numerical simulations with representative parameters for each progenitor and show that the jet succeeds to break out relativistically, which indicates that it potentially create GRBs far away from stellar surface.}
This paper is organized as follows. In Section 2, we describe the
models and methods in this paper. Then, our results will be presented
with detailed analyses in Section 3. Finally, we discuss our
findings and conclude the paper in Section 4.
%In this paper, we employ the geometrical units.
% It should be noted, however, that we also do not discuss and specify
% the mechanism of the jet launch from the central engine. Instead, we
% assume the constant accretion-to-jec conversion efficiency in each
% odel.
\section{Numerical Methods and Models}\label{secnumemethodmodel}
The numerical codes employed in this paper are essentially the same as
those used in \citet{2011ApJ...731...80N}, in which all the details
about our numerical codes and various test calculations are
presented. Here we briefly summarize the methods and setups in this
study.

Our numerical code solves the relativistic hydrodynamic equations with
a weak gravitational field. The self gravity is included in the weak
field approximation of the Einstein equation. It should be noted that, due
to our computational limitations, we cut the inner portions of the
star from a certain radius. The gravity in this region is added as
that of a point mass at the center
%and its mass evolves with time 
by integrating the mass flux at the inner boundary. Based on the above
assumptions, the gravity is solved by using MICCG methods. The
hydrodynamical parts are solved by using the so-called central scheme,
which guarantees good accuracy even if the flow involves strong
shock waves and the flow velocity is highly relativistic
\citep{Kurganov2000,2008ApJ...689..391N}. We use the PPM interpolation
method and TVD Runge-Kutta time integration which achieve second
order accuracy in both space and time. In this study, we adopt the
$\gamma$-law equation of state (EOS), $p = (\gamma - 1) \rho_{0}
\epsilon$ with $p$, $\gamma = 4/3$, $\rho_{0}$ and $\epsilon$ being
the pressure, adiabatic index, rest mass density and specific internal
energy, respectively, in all our computations.

\begin{figure}
\vspace{15mm}
\epsscale{1.0}
\plotone{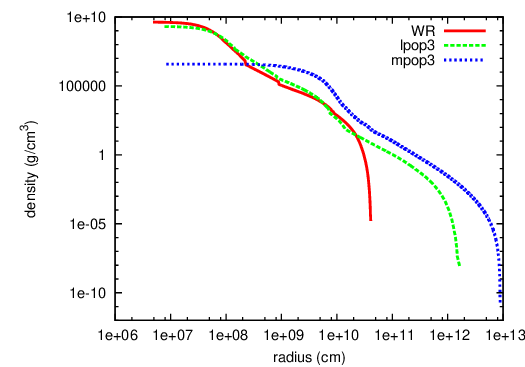}
\caption{The density profiles of the progenitor models used in this
  paper (see Table~\ref{tab1}).
%{\bf ??? why mpop3 has two lines above $10^{11}$ cm?}
\label{f1}}
\end{figure}

In this paper, we adopt three representative stellar progenitors,
which are (1) Wolf-Rayet star (16TI in \citep{2006ApJ...637..914W},
hereafter WR), (2) light Pop III star $40M_{\sun}$, which is the metal
free pre-supernova model calculated by \cite{2002RvMP...74.1015W}
% (??? Woosley et al. 2002 calculated the evolution of the metal free
% star. please mention something???)
 (hereafter lpop3), and (3) massive Pop III star
%\naga{$\sim 1000 M_{\odot}$}
$915M_\odot$
\citep{2009ApJ...706.1184O} (hereafter mpop3).
% The lpop3 progenitor model is employed by the motivated by the {\bf
% Hosokawa et al.}.
The density profile of each stellar model is displayed in
Figure~\ref{f1}, and the stellar mass and radius are
also summarized in Table~\ref{tab1}. Since this study is the first
attempt for the accretion-powered jet propagations, we neglect the
stellar rotation for simplicity although these progenitors are
supposed to spin rapidly to operate the central engine.
 Note that we use the same approach as in the previous study
\citep{2011ApJ...726..107S}, that the accretion-to-jet conversion
efficiency parameter absorbs this uncertainty. More detailed studies
for the effects of rotation will be presented in the forthcoming
paper \citep{2012nagetal}.

We map the spherical symmetric progenitors into two-dimensional grids
in spherical coordinates. The computational domain for each stellar
progenitor covers from a certain inner radius to the stellar
surface. Although the inner boundary should be located in the vicinity
of a black hole (around $10^{6-7} {\rm cm}$), this is computationally
very expensive, so that we set the inner boundary far from a black hole.
For our reference models, the inner boundary for each progenitor is located at $R_{in} \sim 10^{-2} \times R_{star}$, where $R_{star}$ denotes the stellar radius (see Table~\ref{tab2}). According to this limitation, our discussions in the present paper are at the qualitative level.
% Besides, we conduct numerical simulations in lower resolutions than
% previous studies \citep{2009ApJ...700L..47L,2011ApJ...731...80N},
% since it is not necessary to discuss the internal strcture in detail
% as the purpose of this study is concerned.
%\naga{The outer boundary for each models is set in slightly outside from each progenitors. For WR model, it locates at $4 \times 10^{10} {\rm cm}$, while one for mpop3 and lpop3 locate at $1.7 \times 10^{12}  {\rm cm}$ and $10^{13} {\rm cm}$, respectively. }
The outer boundary for each model is set in slightly outside from each progenitor. It locates at $4 \times 10^{10} {\rm cm}$, $1.7 \times 10^{12}  {\rm cm}$ and $10^{13} {\rm cm}$ for WR, lpop3 and mpop3, respectively.
For all models, the number of standard radial grid points is 500. 
The grid width is non-uniform and increasing
%\naga{??? logarithmically}.
in geometric progression \citep{2011ApJ...731...80N}. 
%\naga{??? what is the meaning of geometrically? ???}
%\naga{???logarithmically would be better???}
The innermost grid width is set as $\Delta r = R_{in}/10$ where
$R_{in}$ is the radius of the inner boundary from the center. Then, the rate of geometrical
increase is determined so as to cover all computational regions with
500 meshes. The angular grid covers a quadrant of the meridian section
(where we assume equatorial symmetry) and is uniform with 60 grid
points. We employ an adaptive mesh refinement (AMR) technique in order to decrease the
computational cost. 
%\naga{??? How about the criteria of AMR???}
We deploy two levels of meshes as in
\citet{2011ApJ...731...80N},
%{\bf where} the resolution of {\bf the} second level is finer than
%{\bf the} first level.
where the resolution of the second level is 3 times finer in each
direction than the first standard mesh.
Thus, the angular resolution in AMR region corresponds to $0.5^{\circ}$ for all models. The smallest radial grid width for WR model is $1.6 \times 10^{7} {\rm cm}$, while one for lpop3 and mpop3 are $3.4 \times 10^{8} {\rm cm}$ and $3.4 \times 10^{9} {\rm cm}$, respectively.
 In order to check the
dependence on the resolution, we also carry out finer AMR
calculations.  We check calculations with 5 times and 7 times finer
AMR meshes for WR models (WRreso5 and WRreso7), and 7 times for other progenitor models (lpop3reso7 and mpop3reso7).
For 5 times finer AMR meshes, the angular resolution corresponds to $0.3^{\circ}$. The smallest radial grid widths are $9.6 \times 10^{6}{\rm cm}$ (WR), $2.0 \times 10^{8} {\rm cm}$ (lpop3) and $2.0 \times 10^{9} {\rm cm}$, respectively. For 7 times finer AMR meshes, the angular resolution is $(3/14)^{\circ}$. The smallest radial grid widths are $6.9 \times 10^{6}{\rm cm}$ (WR), $1.5 \times 10^{8} {\rm cm}$ (lpop3) and $1.5 \times 10^{9} {\rm cm}$, respectively. As we shall see in Section~\ref{subseclimitation}, the numerical resolution is important to prevent baryon pollution by numerical diffusion (although it does not affect the macroscopic jet dynamics. See Section~\ref{subseclimitation}).
%We confirm that the resolution does not change our main findings in this
%study.
 Note that in Table~\ref{tab3}, the ``AMR level'' denotes the multiplying
factor of the fine meshes.
%{\bf (??? in Table 3, what is the number of ``AMR level''? second level resolution divided by the first level one? need some explanation???)}

We assume that the central engine successfully operates and the well
confined outflows are produced in the vicinity of a black hole.  We
%demonstrate this situation to
inject the plasma in the radial direction through the inner boundary
with an opening angle of several degrees. In this paper, we also assume
that the jet luminosity depends only on the
%is determined only as a function of 
accretion rate which is estimated by the mass flows across through the
inner boundary;
\begin{eqnarray}
\dot{M} \equiv - 2 \pi \int_{0}^{\pi} \rho_{0}(r_{in},\theta) v^{r} (r_{in},\theta) {r_{in}}^2 \sin{\theta} d{\theta} \label{eq;mdot},
\end{eqnarray}
where $r_{in}$ and $v^{r}$ denote the location of the inner boundary
and the radial velocity of flows. We inject an outflow with a
luminosity,
\begin{eqnarray}
L_{jet} = \eta \dot{M} c^{2}; \label{eq;Ljet}
\end{eqnarray}
where $\eta$ and $c$ denote the accretion-to-jet conversion efficiency
parameter and the speed of light, respectively.
The maximum conversion efficiency as a consequence of the accretion process is $5.7 \%$ for a Schwarzschild black hole and $42 \%$ for an extreme Kerr black hole \citep{1983bhwd.book.....S}. Thus, we choose the value of $\eta$ less than that. The conversion efficiency, jet opening angle, specific
internal energy and radial velocity are varied in each model. Once
these parameters are fixed, the density and pressure of injected jets
are determined by using the relation,
%relations with the power of jet;
\begin{eqnarray}
L_{jet} = \rho_{0} \Gamma v^{r} (h \Gamma - 1) c^2 \Delta{S} \label{eq;Ljetrela}
\end{eqnarray}
where $h (\equiv 1 + \epsilon/c^2 + p/(\rho_{0} c^2))$
%\naga{where $h (\equiv 1 + \epsilon + p/\rho_{0})$}
%\naga{??? expression in natural unit might be better???}
and $\Delta{S}$ denote the specific enthalpy and the area of the
injection surface, respectively.

The collapse of the massive envelope is induced in the same way as in
\citet{2011ApJ...731...80N}. In reality, the stellar envelope begins
to fall after the arrival of a rarefaction wave that is generated by
the inner core collapse. We mimic this situation by putting the radial
gradient of all quantities to zero at the inner boundary except for the jet
injected regions. At the beginning of the simulation, the break of the
force balance at the inner boundary induces the infall of
matter. Subsequently a rarefaction wave propagates outwards, inducing
the infall as it reaches each point. It should be noted, however,
that the stellar progenitors used in this paper, especially the WR progenitor,
are not exactly in dynamical equilibrium, in contrast to
\citet{2011ApJ...731...80N}. Accordingly, the outer parts of the
envelope begin to move and the several artificial waves are observed
during the simulations. We find that they induce artificial explosions
in some models (but only for non-successful shock breakout
models). This is because the WR stellar model, for example,
originally involves rotation \citep[see
  e.g.][]{2006ApJ...637..914W}, so that the centrifugal force works to
sustain the stellar configurations. Since we artificially remove rotation,
 the envelope tends to infall and induce artificial
compressions and bounces. We note that even the original
16TI model is not in exact dynamical equilibrium.
 Contrary to \citet{2011ApJ...731...80N}, we do not take
special treatments for the initial stellar configurations here.

\begin{figure}
\vspace{15mm}
\epsscale{1.0}
\plotone{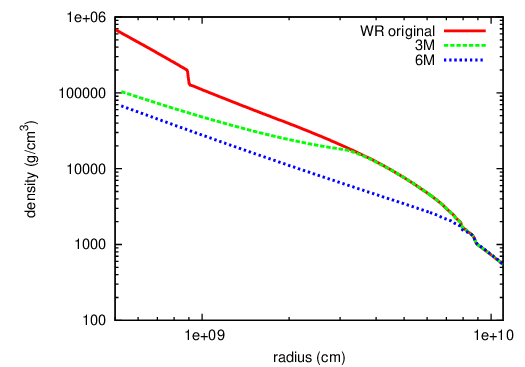}
\caption{The radial density profile for the late time injection
  models. The red line is the original stellar profile of WR model
  while the green (blue) line indicates the density profile at the
  time when the enclosed mass at the inner boundary becomes $M = 3
  M_{\sun}$ ($M=6 M_{\sun}$) as a result of the spherical collapse.
  The rarefaction waves can be clearly seen in this figure.
\label{f2}}
\end{figure}

We also investigate the dependence on the timing of the jet injection,
since we still have few constraint on the starting time of
%operation timing of 
the central engine. 
%Although the centrifugal bounce, which was found
%in the previous studies
%\citep{2010ApJ...713..800L,2011ApJ...731...80N}, is suppressed due to
%the neglect of rotation in this study, 
If the operation of the central engine is sufficiently late, the
density profile of the stellar envelope is changed by the accretion.
Thus, it is expected that the jet dynamics also depends on the timing
of the jet injection.
%the change of stellar configuration. 
In addition, we would like to investigate whether the later jet can
really accomplish the shock breakout since the large mass accretion
may prevent the jet propagations in this case. Motivated by these
facts, we initially let the stellar envelope spherically collapse, and
then inject a relativistic jet. We carry out
these simulations only in WR models
%The reason why we consider only in WR models is that
because the enclosed mass at the inner boundary for our reference
model is $M_{in} \sim 2 M_{\sun}$, which may be lower than the
critical mass of the black hole formation
\citep{2010Natur.467.1081D}. As a result, it is quite likely
that the central engine does not operate for a while.  On the other
hand, the Pop III models have so much mass enclosed at the inner
boundary (see Table~\ref{tab2}) that the central engine would begin to work
soon after the collapse in our models.  We prepare two models, WRM3
and WRM6, which inject the jet when the enclosed mass at the
inner boundary reaches $M_{in} = 3 M_{\sun}$ and $6 M_{\sun}$ ,
respectively. The corresponding retarded time of injection for each model is
$t_{late} = 7.47$s and $26.93$s, respectively. The radial density
profiles just before the jet injection are displayed in
Fig~\ref{f2} for these models.

All of our models used in this paper are summarized in
Table~\ref{tab2}.  We prepare a reference model for each progenitor
with the following parameters.
%We determine {\bf all the} parameters for the reference model as {\bf
%follows}.
The inner boundary is located at $R_{in} \sim 10^{-2} \times R_{star}$,
where $R_{star}$ denotes the stellar radius. The accretion-to-jet
conversion efficiency ($\eta$), half opening angle of outflows
($\theta_{op}$), injected Lorentz factor ($\Gamma_j$), injected
specific internal energy ($\epsilon_j$) are set as $\eta = 10^{-3}$,
$\theta_{op} = 9^{\circ}$, $\Gamma_j = 400$ and $\epsilon_j =
10^{-2}$, respectively. Note that these models assume that the
injected jet has already reached to the terminal Lorentz factor. It
may not be true for WR models since the inner boundary is located
somewhat closer to the black hole than the other progenitor
models. However, as we shall see, if the choice of the terminal
Lorentz factor is the same, the overall profiles of jet dynamics do
not depend on whether the jet injection is kinetic dominant or thermal
dominant
%the kinetic dominant or thermal dominant injections 
(we demonstrate $\Gamma_j = 5, 50$ cases in WRLo5 and WRLo50
models). In order to study the dependence on the accretion-to-jet
conversion efficiency, we vary it as $\eta = 5 \times 10^{-4}, 2
\times 10^{-4},$ and $10^{-4}$ (models such as WRef..., lpop3ef...,
mpop3ef...), while other parameters are identical to the reference
model.
%As well as the study of the conversion efficiency, 
We also study the dependence on the half opening angle with
$\theta_{op} = 3^{\circ}, 6^{\circ}, 18^{\circ}, 36^{\circ},$ and
$45^{\circ}$. The study for the injection timing is done only in WR
progenitor as WRM3 and WRM6 models. As we have already mentioned, we
check the dependence on the location of the inner boundary (models
such as WRin..., lpop3in..., mpop3in...) and we also conduct the
resolution checks (models such as WRreso..., lpop3reso...,
mpop3reso...).

 According to these studies, we find favorable conditions for creating GRBs. In consideration of these results, we also perform numerical simulations with representative parameters (models such as WRrepr, lpop3repr, mpop3repr). As shown in the following section, representative models succeed a powerful relativistic jet breakout and they are the most guaranteed candidates to create GRBs in our models.

%As shown in Section~\ref{subseclimitation}, our current numerical calculations have lots of limitations to determine the possiblity of GRBs and we can not give any strong claims for the possiblity of GRBs. As shown in the following section, representative models succed powerful relativistic jet break out. Thus this fact supports the idea that the massive Pop III stars are progenitor candidates for GRBs.

\section{Result and Analysis}
In this section, we describe the numerical results obtained from our
hydrodynamic simulations and analyze them in detail. We first explain
overall features of the jet dynamics, and then we further analyze the
dependence on each parameter and model.  We also present simple
analytic criteria for the possibility of GRB production
%relativistic jet breakout
at the end.

\subsection{Basic Feature} \label{subsec:basicfeature}
We summarize numerical results in Table~\ref{tab3}. Here, we define
the shock breakout as that
 the forward shock wave successfully reaches
the stellar surface. It should be noted, however, that shock breakout
is a minimal requirement for producing GRBs. As we shall see, even if
the outflow successfully accomplishes the breakout, some models are
not suitable for GRBs (see the column of ``\textit{Possibility of
  GRB production}'' in Table~\ref{tab3}). We assess the possibility of GRB production by
the diagnostic terminal Lorentz factor $\Gamma_{dt} \equiv h \times \Gamma$ profiles on z-axis at the time of the shock breakout.
(Note that the diagnostic terminal Lorentz factor is the achievable Lorentz factor after the internal energy is converted into the kinetic energy.)
If there are regions where $\Gamma_{dt} \ge 100$ is satisfied,
  we determine that the model has a potential to produce GRBs. If this
  condition is not satisfied, the jet is not
  successfully injected or does not move forward. As a result, the
  explosion would never become relativistic enough.
% i.e. the flow is optically thick to gamma-rays without GRBs.
 Note that some models are hard to be judged for the possibility of
  GRB production because of several reasons
 (see the column of ``\textit{Possibility
    of GRB production}'' in Table~\ref{tab3} and we describe these models as
  $\triangle$. See also the discussion in Section~\ref{subseclimitation}).
 The details of these models are presented in the following subsections.
%% If there are regions where $\Gamma_{dt} \ge 100$ is satisfied and the outer envelope has mildy relativistic $\Gamma_{dt} \ge 10$ (i.e., the breakout is enough relativistic),
%%  we decide that the jet has a potential to produce GRBs.
%% \naga{??? If 
%% the diagnostic terminal Lorentz factor
%% is $\Gamma_{dt} \ge 100$ in some region
%% and mildly relativistic $\Gamma_{dt} \ge 10$
%% at the initial surface 
%% ??? when the breakout occurs ???
%% (i.e., the breakout is enough relativistic),
%%  we decide that the jet has a potential to produce GRBs. 
%% ??? what is the meaning of the outer envelope? ???
%% The jet will push sideways the preceding slow outflow,
%% which will form a fast cocoon component around the jet
%% \citep{2011PThPh.126..555I}.}
%If this condition is not
%satisfied, the relativistic outflows can not be expected, even if the
%forward shock wave reaches the stellar surface. We will argue
%further in each subsection.
% As shall we see, even if the outflow successfuly accomplishes break
% out, key elements such as diagnostic terminal Lorentz factors (see
% subsection~\ref{subsecdepeoneff} for the definition of deagnostic
% terminal Lorentz factors) for some models are not suitable for GRBs
% (See the line of \textit{possibility of GRB} in Table~\ref{tab3}. We
% will concern and discuss them further in each subsection.
One of the remarkable results in this study is that many models
successfully accomplish the relativistic shock breakout even though
the progenitor star has an extremely large envelope such as Pop III stars.
%\naga{In fact, numerical simulations for mpop3 with representative parameters actually produce the strong relativistic shock breakout.}
 These results are the numerical verification of \citet{2011ApJ...726..107S}.
 We will further analyze the dependencies
on the efficiency $\eta$ and the opening angle $\theta_{op}$ in the
following section. Indeed, the jet dynamics strongly depend on these
key parameters.
% of the injection conditions.

At the beginning of the simulations, the infall starts at the innermost
regions and subsequently a rarefaction wave propagates outwards. The
accretion rate increases with time and then the injected energy
generates the forward shock wave around the polar regions. It should
be noted, however, that the forward shock wave does not propagate outward
into the active computational regions for a while because of the
interruption by the infalling matter.  If the injected energy is not
enough to push the infalling stellar mantle aside, 
the mantle advects inwards and
eventually it is swallowed into a black hole. As a result, the total
amount of explosion energy becomes less than the injected
energy. Furthermore, the binding energy by a black hole (and also the
progenitor star itself) further reduces the explosion energy. We
also calculate the diagnostic energy $E_{dg}$ at the time of
the shock breakout in each model and show them in 5th rows at
Table~\ref{tab3}.  We define the diagnostic energy $E_{dg}$ as the
integral of $\epsilon_{lc}$ (local energy density), which is the sum
of the internal, kinetic and gravitational energy density, over the
regions with positive $\epsilon_{lc}$ and $v^r$. Note that $E_{dg}$ is
not the isotropic energy but collimation-corrected true energy.
%({\bf ??? energy is isotropic energy? or collimation-corrected true
%energy??})
In the weak gravitational field limit, we can define $\epsilon_{lc}$
as;
\begin{eqnarray}
\epsilon_{lc} = T^{tt} - \rho_{0} \Gamma c^2 + \rho_{0} \psi, \label{eq;Elc}
\end{eqnarray}
where $T^{tt}$ and $\psi$ denote the time-time component of energy
momentum tensor and gravitational potential, respectively.
Note that, for this definition of diagnostic energy density, gravity is treated as an external force. This estimation is valid only when the central core object takes the major role for gravity. If the outer envelope plays an important role for gravitational energy, we need to take into account the self-gravity. So the term of gravitational energy should be contributed as a form of not $\rho_{0} \psi$ but $\frac{1}{2} \rho_{0} \psi$ in Eq.~(\ref{eq;Elc}). However, since we use this value for diagnosing whether the outflow is relativistic or not in this study, the severe definition of this quantity is not necessary. Thus, we use Eq.~(\ref{eq;Elc}) in the present paper.
 As you can see in Table~\ref{tab3}, we find that the diagnostic energy for all
models is less than about a third of the injected energy. Thus, it
implies that the central engine has to produce a larger amount of energy
than that observed as GRBs.

%It should be noted that 
The total amount of explosion energy would be larger than the
diagnostic energy $E_{dg}$ at the breakout since the central engine
%is supposed to 
is able to keep operations after the shock breakout due to the continuing accretion.  Indeed, as we
shall see in the following subsections, some models are active at the
time of the breakout.  Thus, although $E_{dg}$ may be lower than the
typical GRB energy in some models, the outflows would gain
further energy from the central engine to create GRBs after the
breakout.
%far from the stellar surface.

\begin{figure}
\vspace{15mm}
\epsscale{1.0}
\plotone{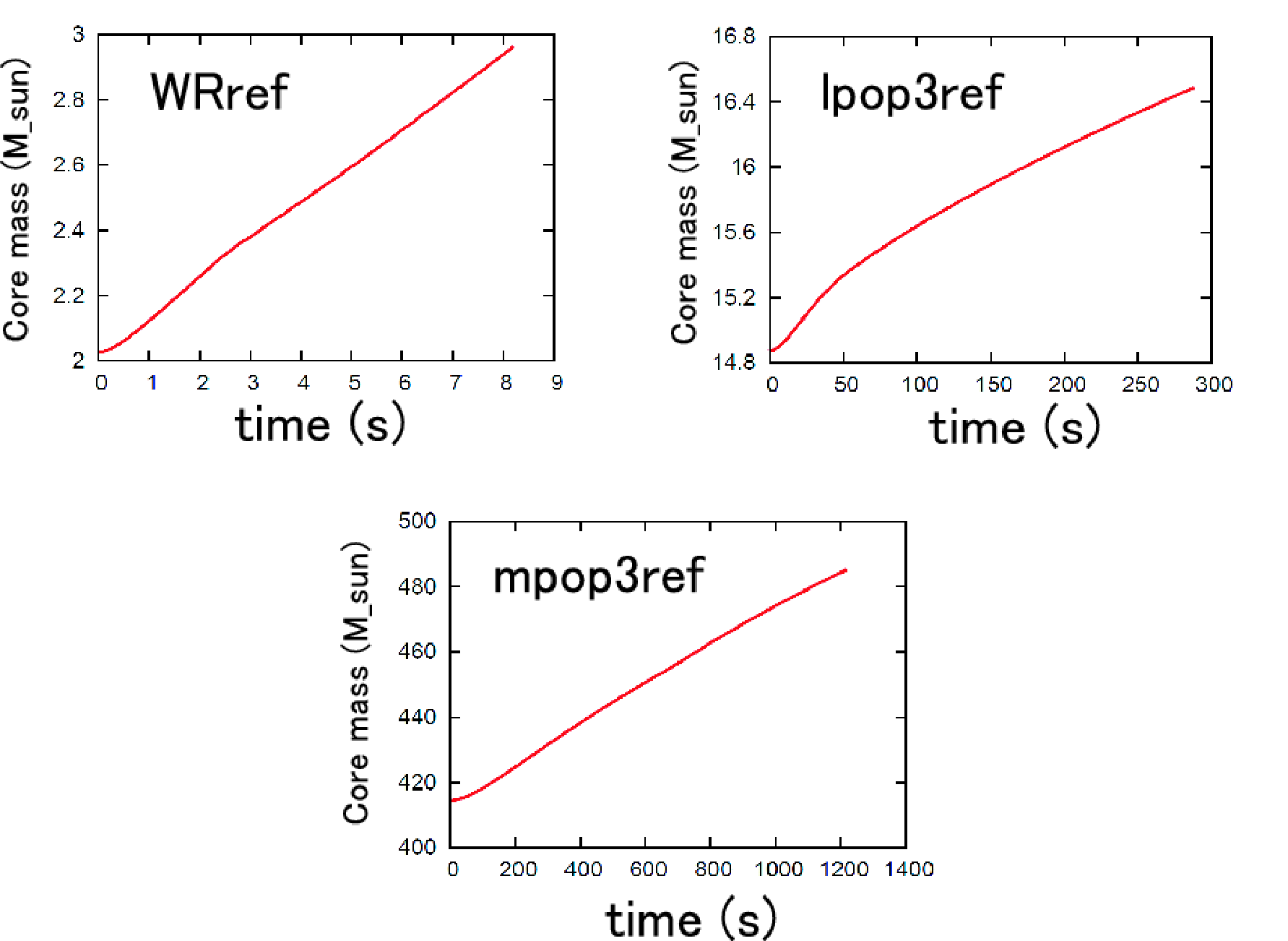}
\caption{The time evolutions of inner core mass (black hole mass)
 for each model.
\label{f3}}
\end{figure}

When the injected energy is successfully launched from the inner
boundary, some portions of matter bounce back and move
outwards. However, we can not see a clear collimated outflow at the
beginning of the simulations even if we inject kinetic dominant
outflows in a small opening angle. This is due to the fact that the
injected energy is thermalized by the strong reverse shock wave in the
vicinity of the inner boundary. As a result, the hot matter expands
and creates a quasi-spherical forward shock wave.  Nevertheless, we
find that the forward shock wave is not strong enough to cease the
infall of matter around the equatorial region and the matter continues to
accrete through the inner boundary 
(see e.g. Fig~\ref{f3} and \ref{f5}).
 The jet structure emerges when
the injected energy becomes enough to push away the reverse shock wave
outside the computational region. Subsequent evolution is similar to
the previous studies of jet propagation. The hot cocoons cause
recollimation shock waves, and strong backflows also appear in the
flows and they sometimes pinch the jet. The Kelvin-Helmholtz instability also
works to create rich internal structures. The jet starts to propagate
and the forward shock wave eventually breaks out of the stellar
surface.

\subsection{Dependence on the Accretion-to-Jet Conversion Efficiency $\eta$} \label{subsecdepeoneff}
\begin{figure}
\vspace{15mm}
\epsscale{1.0}
\plotone{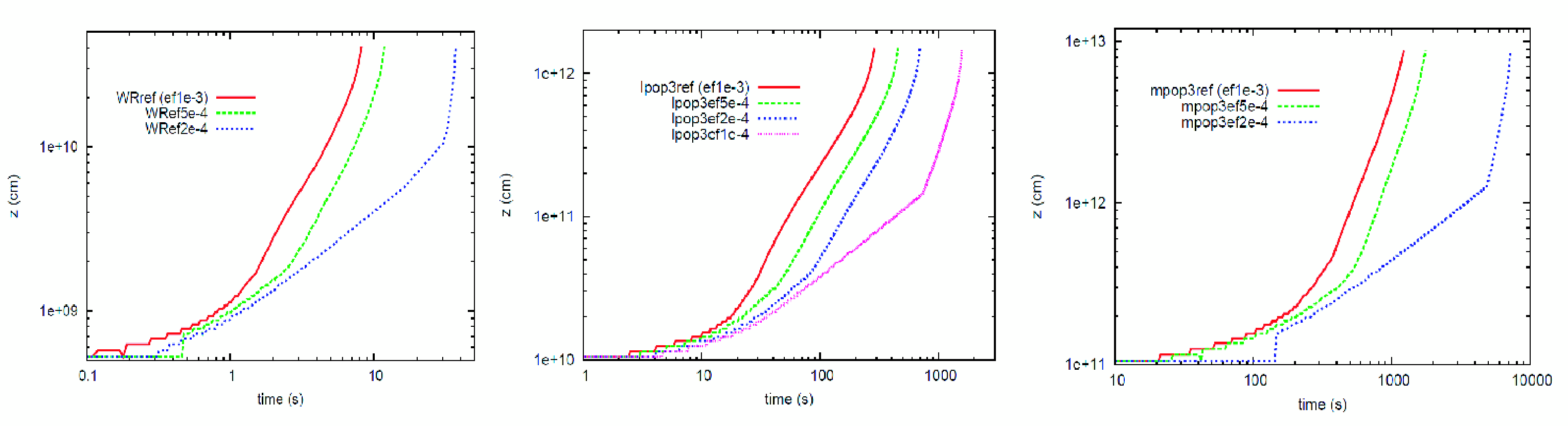}
\caption{ Time evolutions of forward shock waves on z-axis for
  different accretion-to-jet conversion efficiency $\eta$
  models. Left; WR models. Middle; lpop3 models. Right; mpop3
  models. Note that at the beginning of simulations, the artificial
  oscillations arise due to the difficulty of identification of shock position.
 However, they do not affect our main results.
\label{f4}}
\end{figure}

As shown in Table~\ref{tab3}, we find that the forward shock wave can
break out if the accretion-to-jet conversion efficiency is $\eta
\gtrsim 10^{-4}$ (see more details in the subsection \ref{sec:ana}).
%(See also the discussions about the dependence on the inner boundary). 
Figure~\ref{f4} displays the time evolution of the radius for the forward shock wave on the z-axis for models with different efficiencies. As
expected, the higher conversion efficiency generates stronger shock wave,
which quickly propagates into the stellar envelope. In the case of
lower conversion efficiency models, however, it is harder to inject
the jet. Even if the jet is successfully injected, it takes a longer
time for the forward shock wave to reach the stellar surface than in the
high efficiency models.
% It corresponds to the time when the later jet catch up the forward
% shock. It clearly reflects that it takes some time for outflows to
% become the jet structure.

\begin{figure}
\vspace{15mm}
\epsscale{1.0}
\plotone{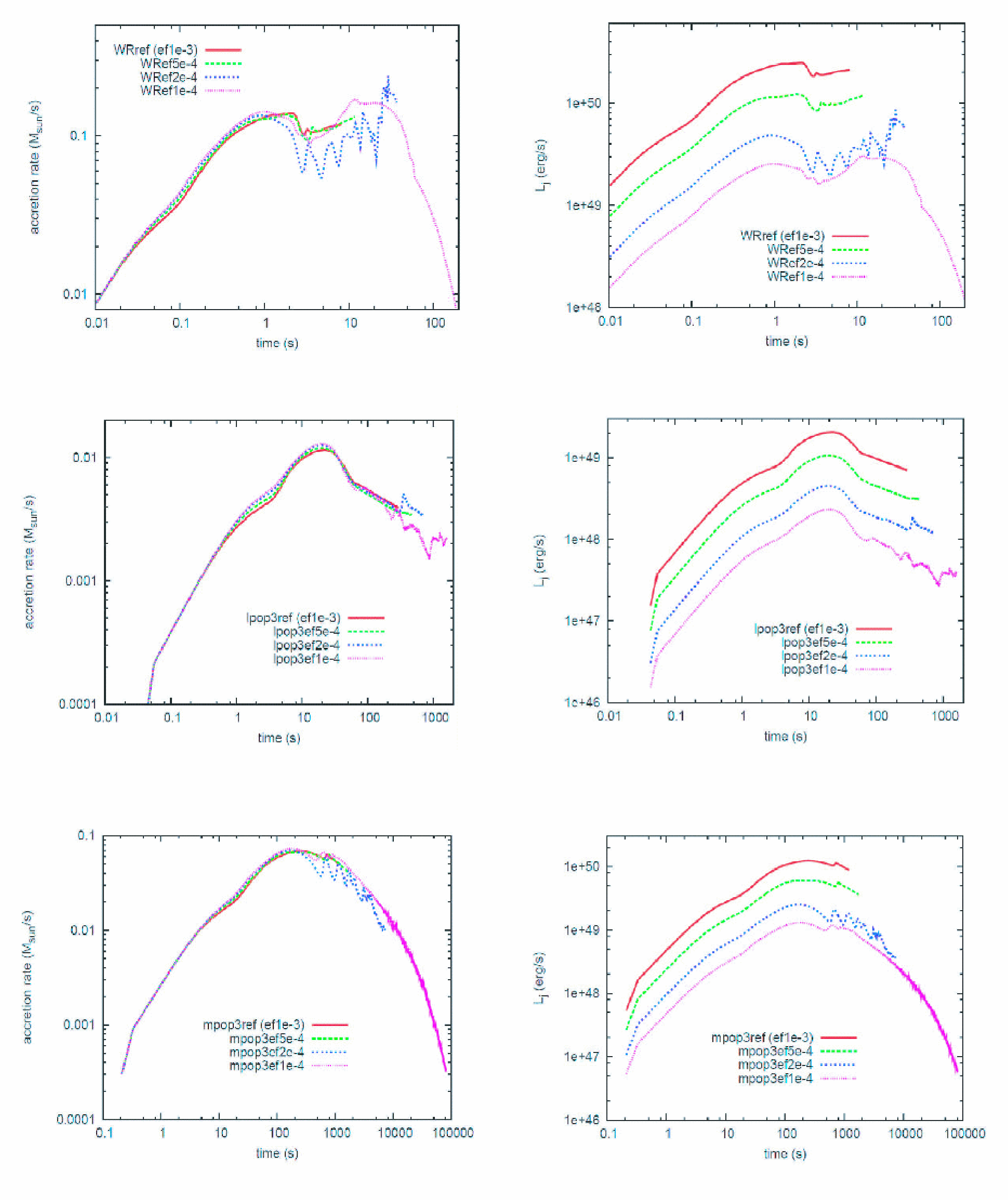}
\caption{The time evolution of accretion rate (left) and luminosity
  (right) for different accretion-to-jet conversion efficiency $\eta$
  models. The upper, middle and bottom panels show WR models, lpop3
  models and mpop3 models, respectively.
\label{f5}}
\end{figure}

Figure~\ref{f5} shows the time evolution of
the accretion rate and luminosity in these models (note that we also
display the results for the failed models without breakouts). At the
initial phase, the higher efficiency models have slightly smaller
accretion rates than the lower efficiency models because
% It is attributed to the fact that 
the strong outflows interrupt the infall of matter. 
%It should be noted that 
Still, the high efficiency models have
%lower accretion rate but 
higher jet luminosity than the low efficiency models.  As a result,
the jet successfully gains energy from the central engine and
accomplishes the shock breakout. It is also interesting to note that
low efficiency models often show rapid time variability of the
accretion rate (see e.g. WRef2e-4 model in the upper left panel in
Figure~\ref{f5}). This is attributed to the
fallback of the shocked envelope, which have rich internal
structures. Although the envelope initially expands due to the
deposited energy, they do not have enough energy to keep moving outwards and
%as a result, 
they eventually fall back through the inner
  boundary, leading to the central engine activity.

Figure~\ref{f6} shows the map of regions with positive
$\epsilon_{lc}$ and $v^r$ at the time of the shock breakout.
%It is immediately seen in this figure 
We can see that the shape of the yellow region, which contains both
positive $\epsilon_{lc}$ and $v^r$ (see Eq.~(\ref{eq;Elc})),
%in outer envelope
does not depend on the accretion-to-jet conversion efficiency $\eta$
so much.  Irrespective of $\eta$, the maximum transverse radii of the
yellow region from the z-axis are $\sim 5 \times 10^{9}{\rm cm}$, $\sim 2
\times 10^{11}{\rm cm}$ and $\sim 1.5 \times 10^{12}{\rm cm}$ for WR,
lpop3 and mpop3 models, respectively.
% ({\bf ??? why? what do you mean by ``does not depends''? at least
% the head velocity is changed. Do you mean the shape of the jet head?
% ???}.
It may imply that the outflow structure 
does not mainly depend on 
%the initial opening angle $\theta_{op}$, not 
the efficiency $\eta$, as long as
the shock breakout occurs (see also the next subsection).
%\naga{We
%  also suggest from this figure that even for lower conversion
%  efficiency case, the jet never become spherical if the initial
%  opening angle of jet is sufficiently small.}
%{\bf ??? if eta is too small, the jet becomes spherical? Why do you
%say that it may not change? ???}
%(Note that, as we shall see in the next subsection, the outflow region
%in the outer envelope strongly depends on the opening angle.).  
From this figure, we can also confirm that even the large efficiency
jet does not expel all portions of the stellar mantle, and matter
can continue to accrete onto the black hole.  On the other hand, the inner
parts of the envelope profiles depend on the efficiency $\eta$.  As
expected, a larger amount of matter is captured for the lower
conversion efficiency.
%leads to  to be captured.  
In addition, as discussed above, the fallback of matter causes the
late time variability of the central engine.

%\naga{As we shall see later in Section~\ref{subseclimitation}, these results are affected by the artificial baryon pollution and the dense ``plug'' at the head of jet due to two-dimensional axisymmetricity of our numerical simulations.}

\begin{figure}
\vspace{15mm}
\epsscale{1.0}
\plotone{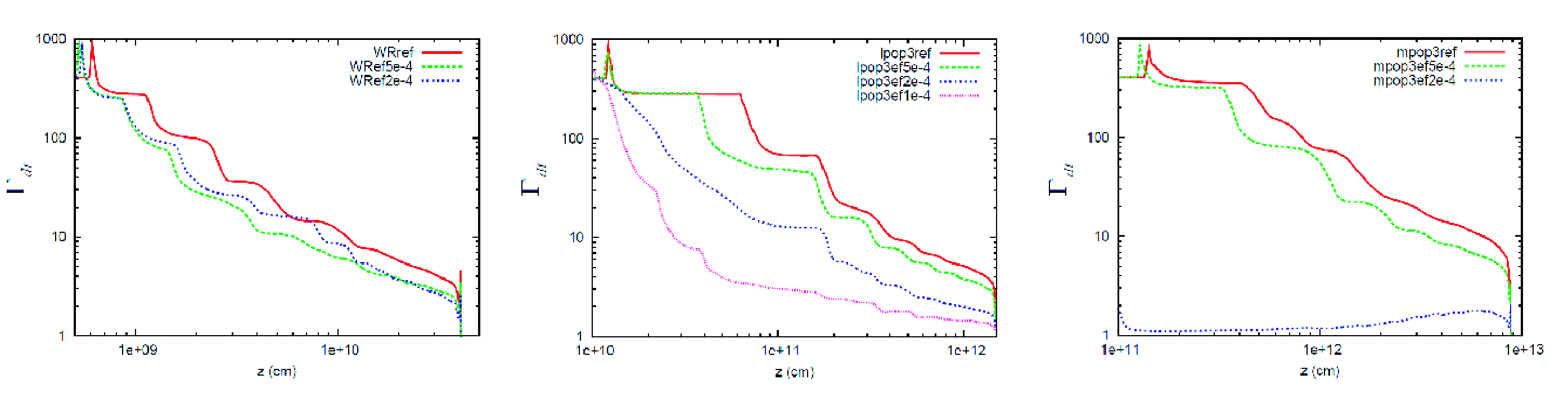}
\caption{The diagnostic terminal Lorentz factor profile ($\Gamma_{dt} \equiv h \times \Gamma$, i.e., the Lorentz factor which the jet can in principle attain) at the time of
  the shock breakout for different accretion-to-jet conversion
  efficiency $\eta$ models. From left to right, WR models, lpop3
  models and mpop3 models, respectively.
\label{f7}}
\end{figure}

In Figure~\ref{f7}, we show
$\Gamma_{dt}$ profiles along z-axis
at the time of the shock breakout for each model.
The model has the potential to produce a GRB if
$\Gamma_{dt} \ge 100$ is satisfied in some region.
Therefore we use this condition to assess whether the model can create GRB or not.
We note that the reality is more complex as argued in the following.
In Figure~\ref{f7},
the outer region tends to contain lower values of $\Gamma_{dt}$,
even less than $\sim 10$.
The small $\Gamma_{dt}$ is caused by the baryon pollution in the jet, most likely because of the lack of the numerical resolution (see in Section~\ref{subseclimitation}). However, even if the baryon pollution were the real physical phenomenon, these outflows would create GRBs for the high 
efficiency models ($\eta = 10^{-3}$) 
because the central engine keeps operating
after the shock breakout for these models (see discussions in Section~\ref{subseclimitation}).
The inner fast moving jets
would eventually catch up the outer slowly moving ejecta. 
Due to this energy input from the inner jets, 
it is quite likely that the actual terminal Lorentz factor 
of the outer outflows is larger than the
values of the current estimation.
% of the diagnostic terminal Lorentz
%factor. 
Thus we expect relativistic explosions in these models.
%\naga{??? what is the final Lorentz factor? can we estimate it? 
%for example, if we separate the jet into the inner and outer parts 
%we can estimate the final Lorentz factor from the two shell models. 
%Or if the inner part has a dominant energy, we can consider
%that the Lorentz factor is determined by this matter. ???}
%{\bf ??? How do you estimate the acceleration? Is the slow jet really
% accelerated? ???}  It should be noted, however, the diagnostic
% terminal Lorentz factor is about less than 10 in the outer
% envelope. Thus, although these models successfully accomplish jet
% breakouts, it it unclear they can really produce GRBs. It should be
% noted, however, that the central engine keeps to operate after jet
% break out, thus, the continuous energy injection will make them
% accelerate. Thus, it has a possibility for high efficiency jet to
% produce GRBs.
%On the other hand, the low efficiency models such as lpop3ef-4,
%lpop3ef2-4 and mpop3ef2-4 are clearly not suitable for
%producing GRBs.
 However, it is less likely that
  the low efficiency models such as lpop3ef-4, lpop3ef2-4
  and mpop3ef2-4 can successfully gain energy from later jets.
  Even if the central engine keeps operating for a long time, they may
  not be able to accelerate outflows because
  the amount of the outer slow matter is large 
(the profile of $\Gamma_{dt}$ deviates from the high efficiency models)
in these models.
%  also keep to prevent the energy injection.
  As a result, the energy
  of jet is dissipated in the vicinity of the injection region 
  %the inner parts of ejecta 
  without
  transferring the energy into the outer parts of the ejecta.
%and it is supposed to be hard to transfer large amount of
%  energy into outer parts of ejecta. 
  According to this
  consideration,
%  speculation,
  we regard that these low efficiency
models would end up with non-relativistic explosions and may not
create GRBs (thus, we mark triangles ($\triangle$) for these models
in the column ``\textit{possibility of GRB}'' of Table~\ref{tab3}.).
%\naga{??? Should we include discussion about the reverse shock evolusion???}
% {\bf ??? How do you judge the impossibility? What is the difference
% between the previous case and this case? ???}
We note that if the reverse shock is stalled
near the injection region at the time of the shock breakout,
there is no hope for this reverse shock to go ahead afterward
because the accretion rate is decreasing.

%Thus, in this case, we will observe them as failed GRBs (See the
%column of ``\textit{possibility of GRB}'' in Table~\ref{tab3}).

\subsection{Dependence on Opening Angle $\theta_{op}$}
\begin{figure}
\vspace{15mm}
\epsscale{1.0}
\plotone{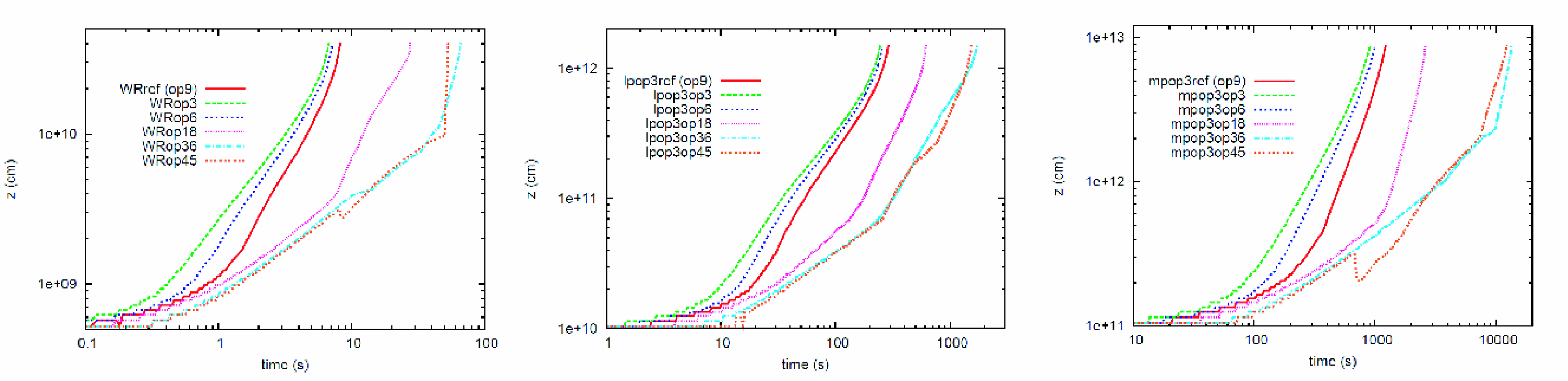}
\caption{ Time evolutions of forward shock waves on z-axis for
  different opening angle $\theta_{op}$ models. From left to right, we
  show WR models, lpop3 models and mpop3 models, respectively.
\label{f8}}
\end{figure}

\begin{figure}
\vspace{15mm}
\epsscale{1.0}
\plotone{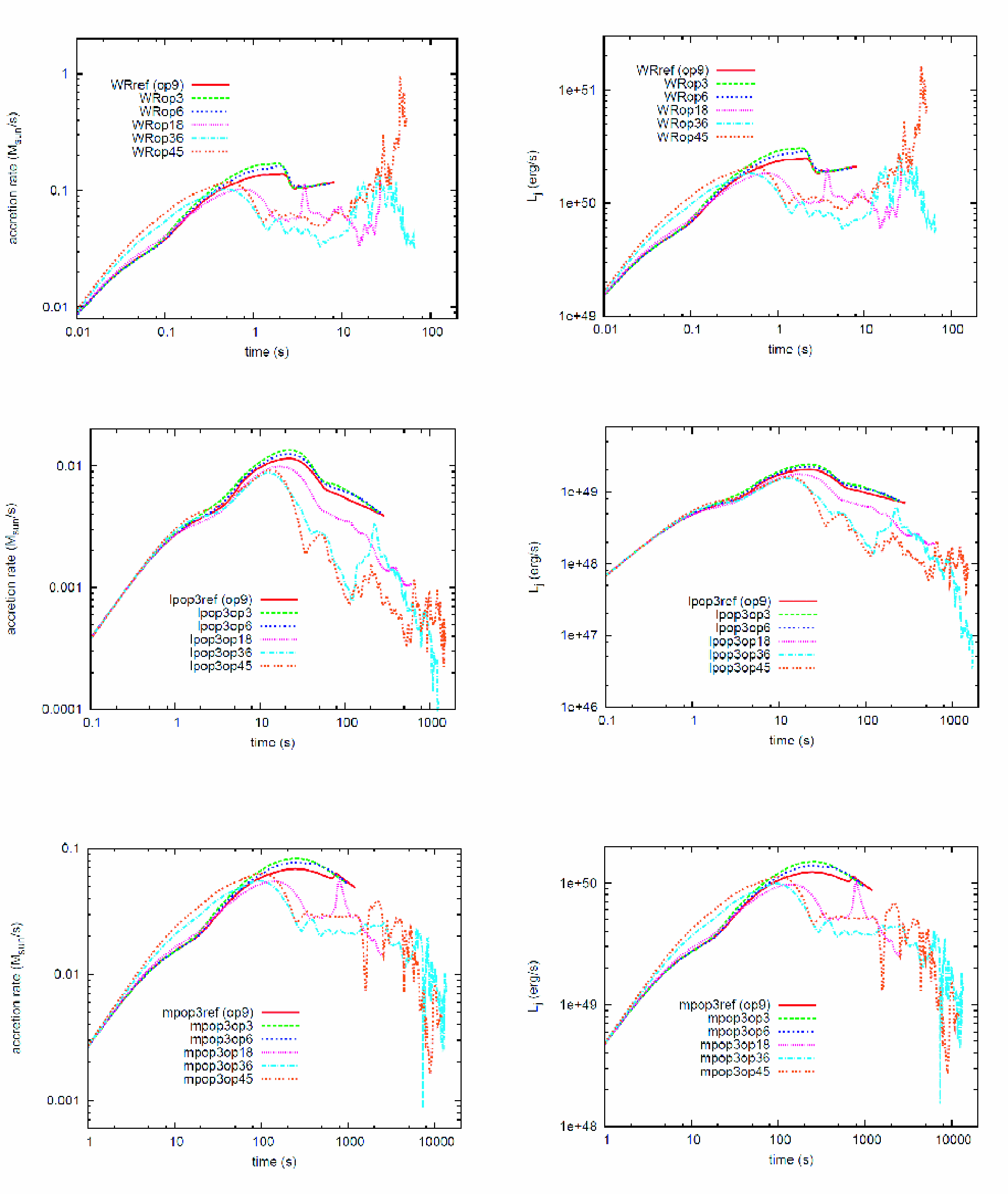}
\caption{The time evolution of the accretion rate (left) and the
  luminosity (right) for different opening angle $\theta_{op}$
  models. The upper, middle and bottom panels show WR models, lpop3
  models and mpop3 models, respectively.
\label{f9}}
\end{figure}

Figure~\ref{f8} shows the time evolution of the 
forward shock wave on the z-axis for models with different opening angles
of the jet at the injection site.
 Generally, the forward shock wave propagates fast and
succeeds in breaking out if the opening angle is small.  It is mainly
attributed to two reasons: One reason is that the isotropic flux is
increased by the small cross-sectional area of the injected jet, and
the other is the increment of the accretion rate.  A wide opening
angle tends to interrupt the accretions and reduce the jet luminosity (see
Figure~\ref{f9}).

Note that the overall dynamics for very wide opening angle jets
(e.g. $\theta_{op} = 36$ and $45^{\circ}$) are more complex than the
collimated jets. As shown in Figure~\ref{f8}, the jet
breakout time for $\theta_{op}=45^{\circ}$ models are (slightly)
earlier than $\theta_{op}=36^{\circ}$ models in all types of
progenitors. It is because the fall back process plays an important
role for wider jets by enhancing the accretion rate.
%drastically. 
Indeed, for WRop45, the accretion rate reaches $\dot{M} \sim 1
M_{\sun}/s$ at $t = 45$ s, and leads to the strong outflows from the
inner boundary and finally to the shock breakout.  Thus, the late time
activity of the central engine is possible for a wide opening angle
due to the strong fallback accretions, even though the mean accretion
rate decreases with time in the late phase.  (But these wide opening
angle jets are not suitable to produce the relativistic outflows, see
below).

% we might expect that the rapid increase of accretion rate due to
% fall back of inhomogeneous shocked mantle causes the late time
% activity of central engine.

\begin{figure}
\vspace{15mm}
\epsscale{1.0}
\plotone{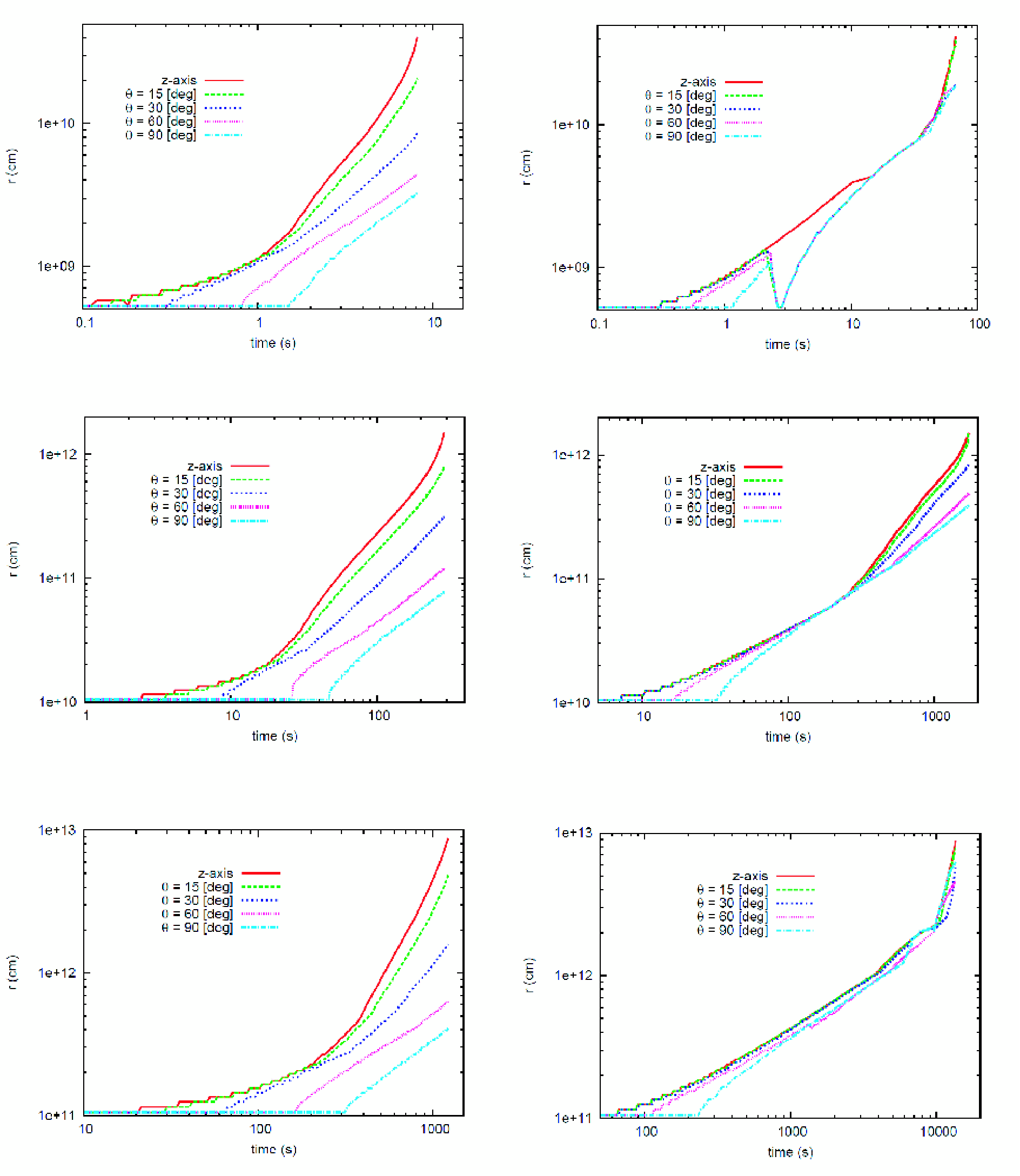}
\caption{The forward shock evolution along each radial ray with
  different angle from the axisymmetric axis.  We show two different
  opening angle jet models. Left: reference model
  ($\theta_{op}=9^{\circ}$), Right: ($\theta_{op}=36^{\circ}$
  model). From upper to lower panels, we show WR, lpop3 and mpop3
  models, respectively.
\label{f10}}
\end{figure}

In Figure~\ref{f10}, we display the time
evolution of the forward shock wave for different angle radial
rays from the symmetry axis.  For reference models (left panels in
this figure), the forward shock velocity decreases with increasing
angle. Particularly, at the time of the shock breakout, we find that
the forward shock waves along the off-axis radial ray are still deep
inside the star, which means that the outflow is well collimated. On
the other hand, for $\theta_{op}=36^{\circ}$ models, almost all models
experience the quasi-spherical evolutions. Although the shock wave on
z-axis is slightly faster than the off-axis shock waves, it is much
more spherical than the reference model at the breakout. As a result,
it is expected that the wide opening angle models never create GRBs
in view of the spherical morphology of the outflows.

\begin{figure}
\vspace{15mm}
\epsscale{1.0}
\plotone{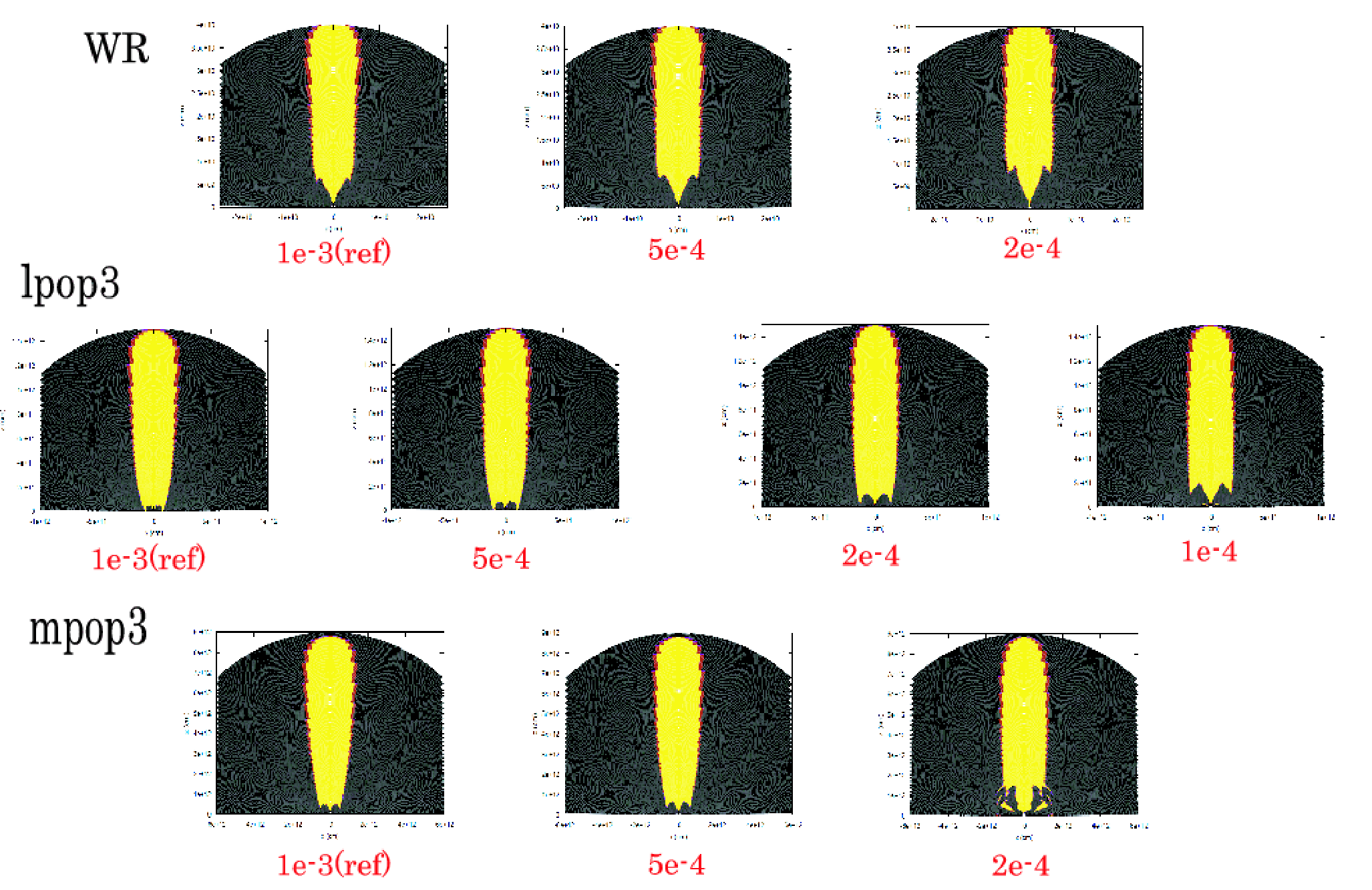}
\caption{The regions with positive $\epsilon_{lc}$ and $v^r$ (Yellow)
  at the time of the shock breakout for different accretion-to-jet
  conversion efficiency $\eta$ models (see section~\ref{subsec:basicfeature} for the definition of $\epsilon_{lc}$ and $v^r$). The upper, middle and bottom
  panels show WR models, lpop3 models and mpop3 models, respectively.
\label{f6}}
\end{figure}

\begin{figure}
\vspace{15mm}
\epsscale{1.0}
\plotone{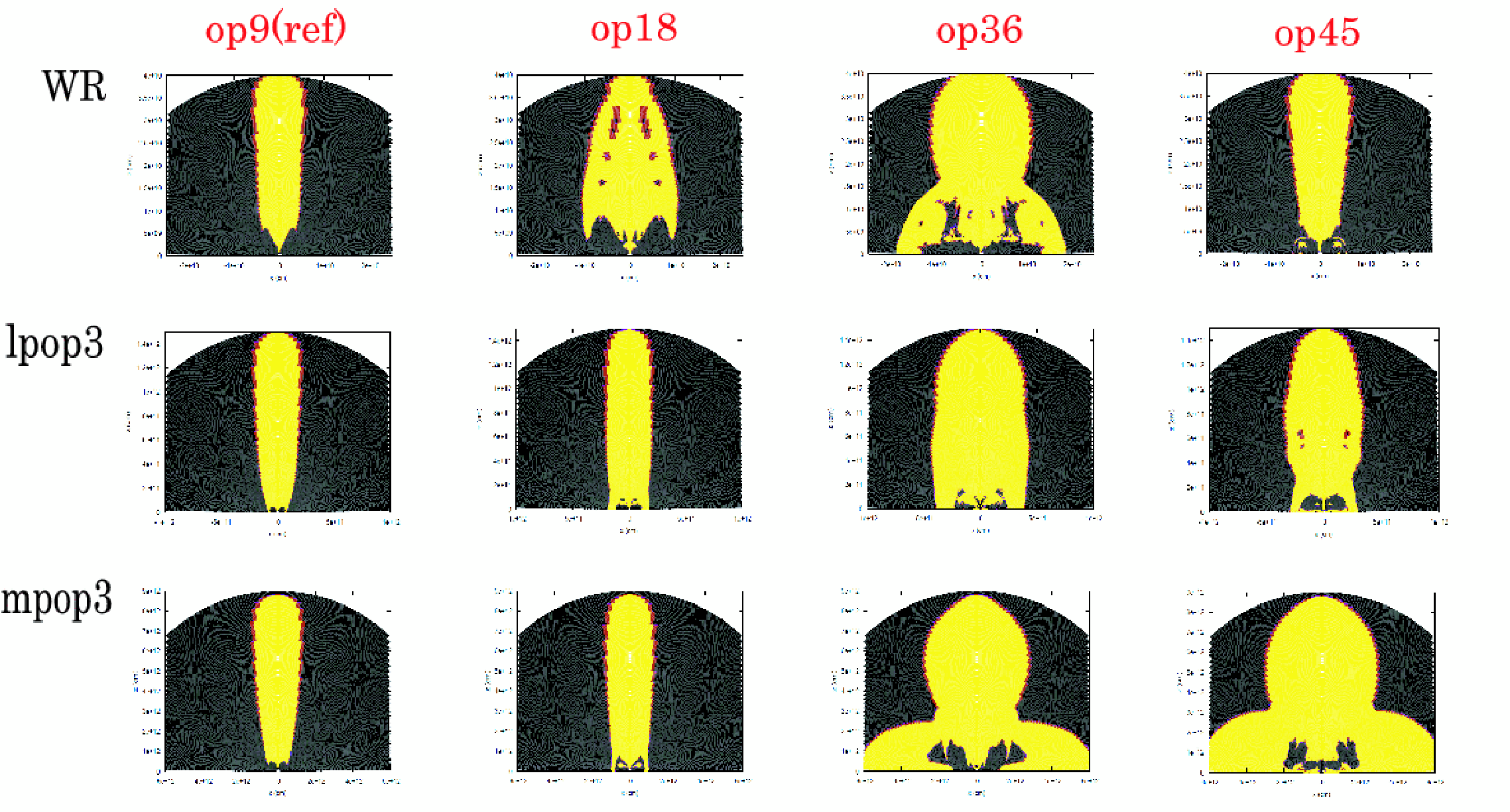}
\caption{ Same as the Figure~\ref{f6} but  for different opening angles.
% The regions with positive $\epsilon_{lc}$ and $v^r$ (Yellow)
%  at the time of the shock breakout for different opening angle
%  $\theta_{op}$ models.
% The upper, middle and bottom panels show WR
%  models, lpop3 models and mpop3 models, respectively.
\label{f11}}
\end{figure}

Figure~\ref{f11} is the same as
Figure~\ref{f6}, but for models with different opening angles. We
do not find any clear differences for narrow opening angle jet models
($\theta_{op} = 3^{\circ}$ and $6^{\circ}$), so we do not display these
models in this figure. As shown in this figure, the wider jet tends to
expel larger outer envelope, and clearly these profiles are different
from each other. Also we again see that very wide angle cases are very
complicated. Interestingly, some fraction of stellar mantle around the
equatorial region may never fall back to the black hole (see
$\theta_{op} \ge 36^{\circ}$ models). From what has been discussed
above, we can conclude that the outflow profile is mainly determined
by the opening angle of the jet $\theta_{op}$ rather than the
accretion-to-jet conversion efficiency $\eta$ (See also
Figure~\ref{f6} and discussions in
subsection~\ref{subsecdepeoneff}).

% ??? if the efficiency is low, the jet becomes spherical? what do you
% mean? ???}

\begin{figure}
\vspace{15mm}
\epsscale{1.0}
\plotone{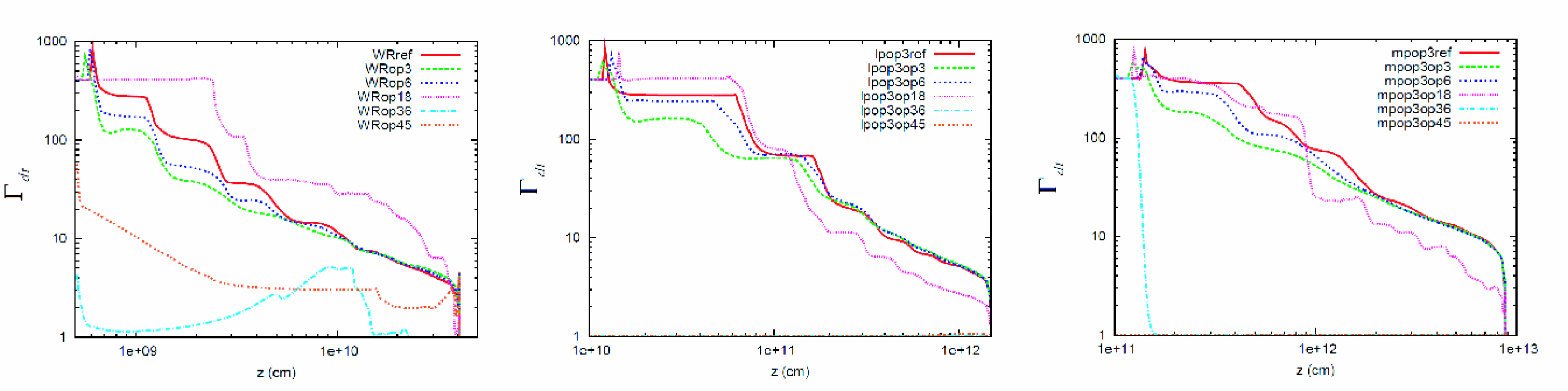}
\caption{Same as the Figure~\ref{f7} but for different opening angles.
%The diagnostic terminal Lorentz factor profile at the time of
%  shock breakout for different opening angle ($\theta_{op}$)
%  models. From left to right, WR models, lpop3 models and mpop3
%  models.
\label{f12}}
\end{figure}

The diagnostic terminal Lorentz factor profiles along the z-axis for
different opening angle models are shown in
Figure~\ref{f12}.  For small opening angle models
($\theta_{op} \le 9^{\circ}$), the diagnostic terminal Lorentz factor
is not different very much (slightly larger value for wider opening
angle). However, the profile of very wide models are completely
different from narrow jet cases and the terminal Lorentz factor for
the wide angle models are quite low. Thus, they are no longer capable
of producing GRBs. Note that, although the jet of mpop3op36
  model is successfully injected around the inner boundary (see in
  Figure~\ref{f12}), this outflow may not create
  GRBs since the outer ejecta is completely non-relativistic
 $\Gamma_{dt} \sim 1$
 and the outflow configuration is not collimated.
%\naga{??? The criteria for the GRBs is the same before? i.e.
%the diagonostic Lorentz factor is larger than 10 in the inner jet? ???}
 As a result, a small opening angle
%\naga{???
%  $\theta_{op} \lesssim 20^{\circ}$ ??? no meaning for the second
%  digit? ???} 
$\theta_{op}\lesssim 20^{\circ}$ is necessary for the relativistic
 shock breakout and GRBs with the case $\eta\sim 10^{-3}$.
%It should be 
Note that if the efficiency becomes lower than $\eta = 10^{-3}$, the
opening angle needs to be smaller than the current value, 
because it is harder for the low efficiency jet to push aside the infall matter than the high efficiency jet (see \S \ref{sec:ana} for the analytic estimation
of the breakout criteria).

\subsection{Dependence on Injection Lorentz factor $\Gamma_j$ and Injection Timing $t_{late}$} \label{subinjeclorentzandtiming}
\begin{figure}
\vspace{15mm}
\epsscale{1.0}
\plotone{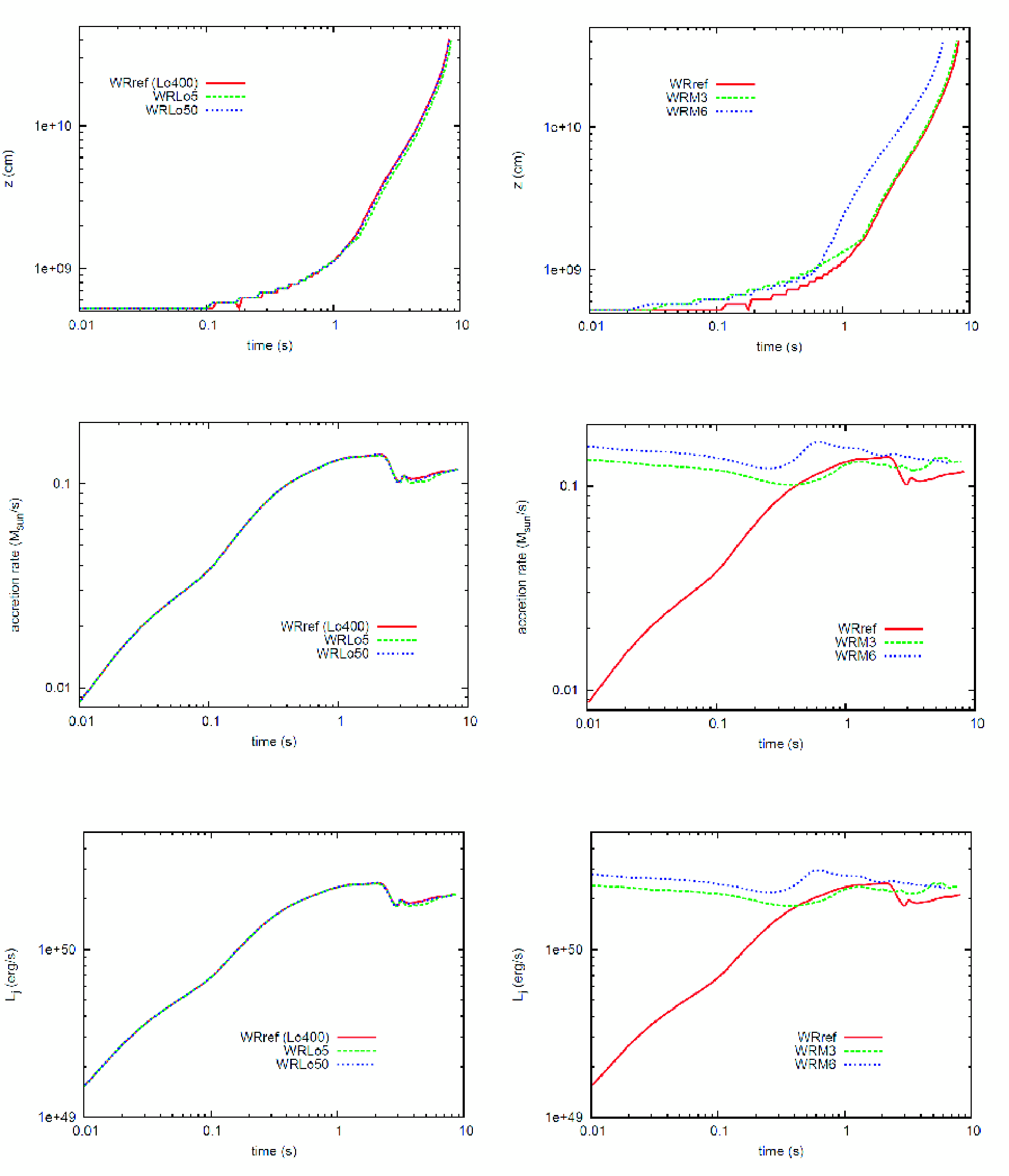}
\caption{The dependence on the injection Lorentz factor and the timing
  of the jet injection. Left: different injection Lorentz factor, Right:
  different injection timing. Upper to lower; the time evolutions for
  the shock evolution along the z-axis, the mass accretion rate and
  the jet luminosity, respectively.
\label{f13}}
\end{figure}

In this section, we discuss how the injection Lorentz factor and the
jet injection timing affect the evolution of the jet. In
Figure~\ref{f13}, we display time evolutions of
some key quantities, such as the forward shock wave on the z-axis, the
mass accretion rate and the jet luminosity. As we can see,
the injection Lorentz factor does not change the qualitative feature
of the jet evolutions for all quantities as long as the final coasting
Lorentz factor is the same (See left panels in
Figure~\ref{f13}).

On the other hand, the jet dynamics depend on the timing of the jet
injection (See right panels in Figure~\ref{f13}).
We can see that the late injection leads to a slightly faster evolution
than the early injection, and also that the time evolution of the
accretion rate and the jet luminosity are quite different in the early
phase.  This is because the accretion rate is already high at the time
of the jet injection for the late injection.  As a result, the strong
jet is injected and evolves faster than the early injection case.  It
should be noted that although the late jet injection slightly
interrupts the infall of the material (see middle and right panel of
Figure~\ref{f13}), it does not cease the accretion
of all the infalling material.  We also speculate that a wide opening
angle would suppress the infall of matter more strongly, so that the
collimated outflows are preferred for GRBs even in the late injection
cases.

\begin{figure}
\vspace{15mm}
\epsscale{1.0}
\plotone{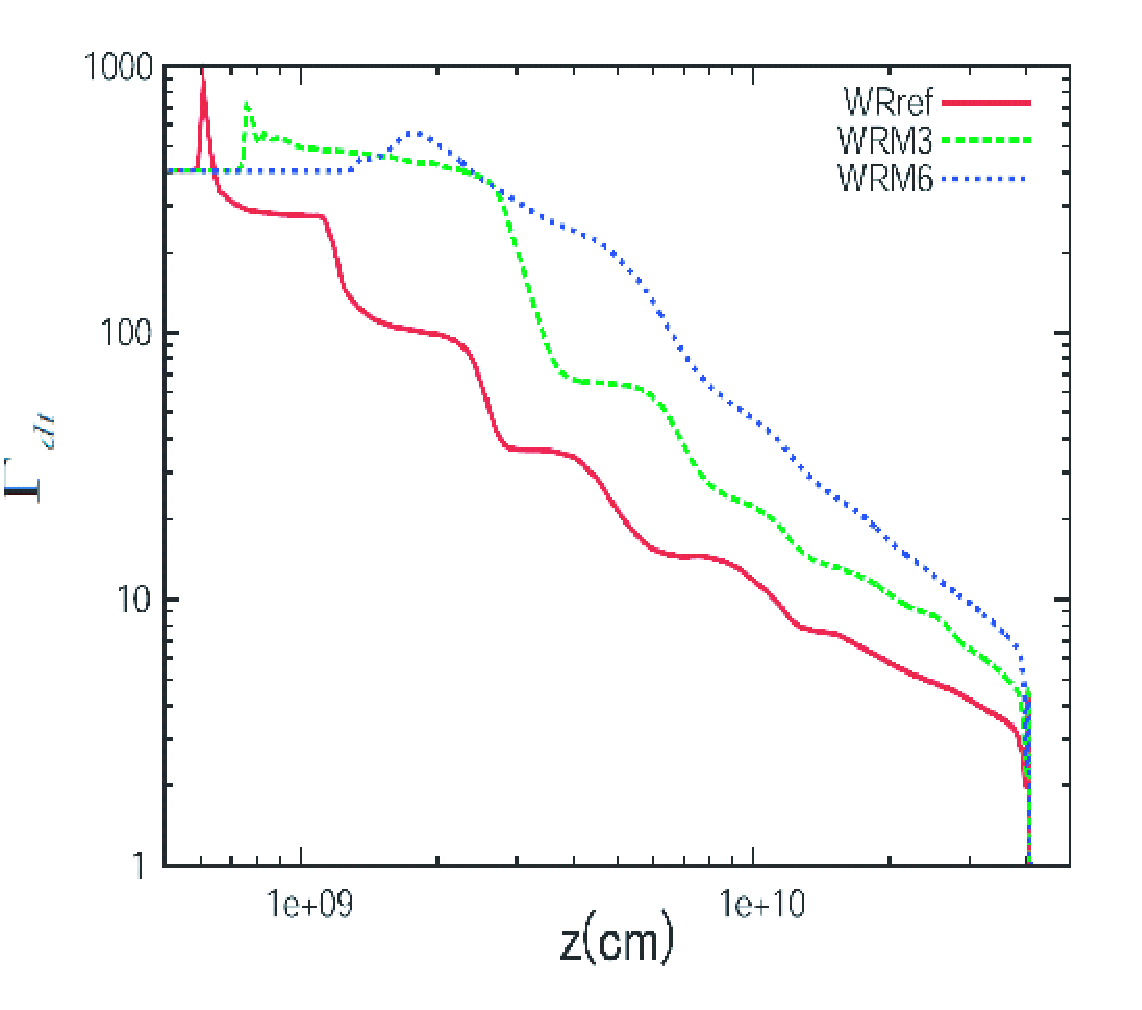}
\caption{Same as Figure~\ref{f7}, but with a
  different timing of the jet injection.
\label{f14}}
\end{figure}

It is also interesting to investigate the diagnostic terminal Lorentz
factor $\Gamma_{dt}$ profile for the different timings of
jet injection as displayed in
Figure~\ref{f14}.  As we can see in
this figure, the late jet injection tends to have higher diagnostic
terminal Lorentz factor than the early jet injection.
%It seems to be attributed to the fact that
 This is also because the jet luminosity for the later injection case is stronger than 
the earlier jet (see bottom panel of
  Figure~\ref{f13}). The strong forward shock waves
  propagate outwards and the outer envelope obtains large energy from
  them.
Besides, we also find that the late jet injection tends to have a smaller amount of baryon mass in the jet since a portion of envelope matter has already been swallowed in the black hole. As a result, the outflow easily achieve a relativistic velocity after the breakout (see Table~\ref{tab4} and Section~\ref{subseclimitation} for more details).
We again must caution that the artificial baryon pollution affects the distribution of diagnostic terminal Lorentz factor and the amount of baryon mass. The detailed discussions of these issues are described in Section~\ref{subseclimitation}.
% and the envelope density is lower at the later time. \naga{Hence} the jet can proceed without affecting the accretion, i.e., the negative feedback is small.
%This is because the envelope density is lower at the later time and
%hence the jet can proceed without affecting the accretion, i.e., the
%negative feedback is small.
%the later jet constantly keep large energy injection while the power
%of outflows of early timing (especially for the WRref model) is weak
%at the beginning of jet injection.  As a result, we now
%Therefore we can conclude that the later jet injection is more
%suitable for producing GRBs than the earlier one.
%\naga{It should be noted that, as we shall see in Section~\ref{subseclimitation}, the artificial baryon pollution may affect the diagnostic terminal Lorentz factor and the amount of baryon mass. Therefore, the obtained diagnostic terminal Lorentz factor is a lower limit and the jet would break out more relativistically.
Note also that too late an injection may not be suitable for
GRBs, since the accretion rate becomes very low and a centrifugal bounce takes place at the later time
\citep[see][]{2011ApJ...731...80N}.  Although we do not know exactly the
starting time of the central engine, we speculate that the most
plausible starting timing for the central engine is the time when the
accretion disk is formed around a black hole.  It clearly depends on
the angular momentum distribution of the progenitor star, so we will
investigate this dependence by performing the rotational collapse of
progenitor stars in the forthcoming paper \citep{2012nagetal}.

% it is the most plausible timing that the accretion disk is
%formed around a black hole. 
%  Thus, at least for
%Wolf Rayet progenitors, the best condition for producing GRBs seems
%that the engine starts when the black hole mass becomes $3 M_{\sun}
%\lesssim M \lesssim 6 M_{\sun}$.
%\naga{??? can you conclude that this mass range is the best?
%Only based on Fig 13? 
%How do you derive the upper and lower mass? ???}
%it seems to be the best condition for producing GRBs which the
%central engine starts to operate at the time when the black hole mass
%become
%We also speculate that the late time jet injection is also better for
%Pop III stars, although we do not have quantitative data.
%for Pop III stars to produce GRBs although we do not know exactly the
%timing of jet formation.
%Note that although we do not know the starting time of the central
%\naga{Although we do not know the starting time of the central
%engine, it is the most plausible timing that the accretion disk is
%formed around a black hole. It clearly depends on the angular momentum
%distribution of the progenitor star, so we will investigate this
%dependence by performing the rotational collapse of progenitor stars
%in the forthcoming paper \citep{2012nagetal}.}

\subsection{Representative Models} \label{subsecrepresent}
Based on the above results, we construct representative models, which are expected to create GRBs in each progenitor. As we have seen, the favorable conditions for the relativistic shock breakout are the high conversion efficiency, small opening angle and late jet injection. We summarize these parameters of representative models in Table~\ref{tab2}. We denote these representative models as WRrepr for the Wolf-Rayet progenitor, lpop3repr for the light Pop III star and mpop3repr for the massive Pop III star, respectively. The accretion-to-jet conversion efficiency is set as $\eta = 10^{-2}$ which corresponds to nearly the maximum conversion efficiency for a Schwarzschild black hole (See Section~\ref{secnumemethodmodel}). The half opening angle is set as $\theta_{op} = 9^{\circ}$. We start to inject the jet when the mass accretion rate becomes the largest for each progenitor. We also note that we carry out simulations with higher spatial resolutions (7-level AMR) in order to suppress the artificial baryon pollutions (See in Section~\ref{subseclimitation} for more details.).

\begin{figure}
\vspace{15mm}
\epsscale{1.0}
\plotone{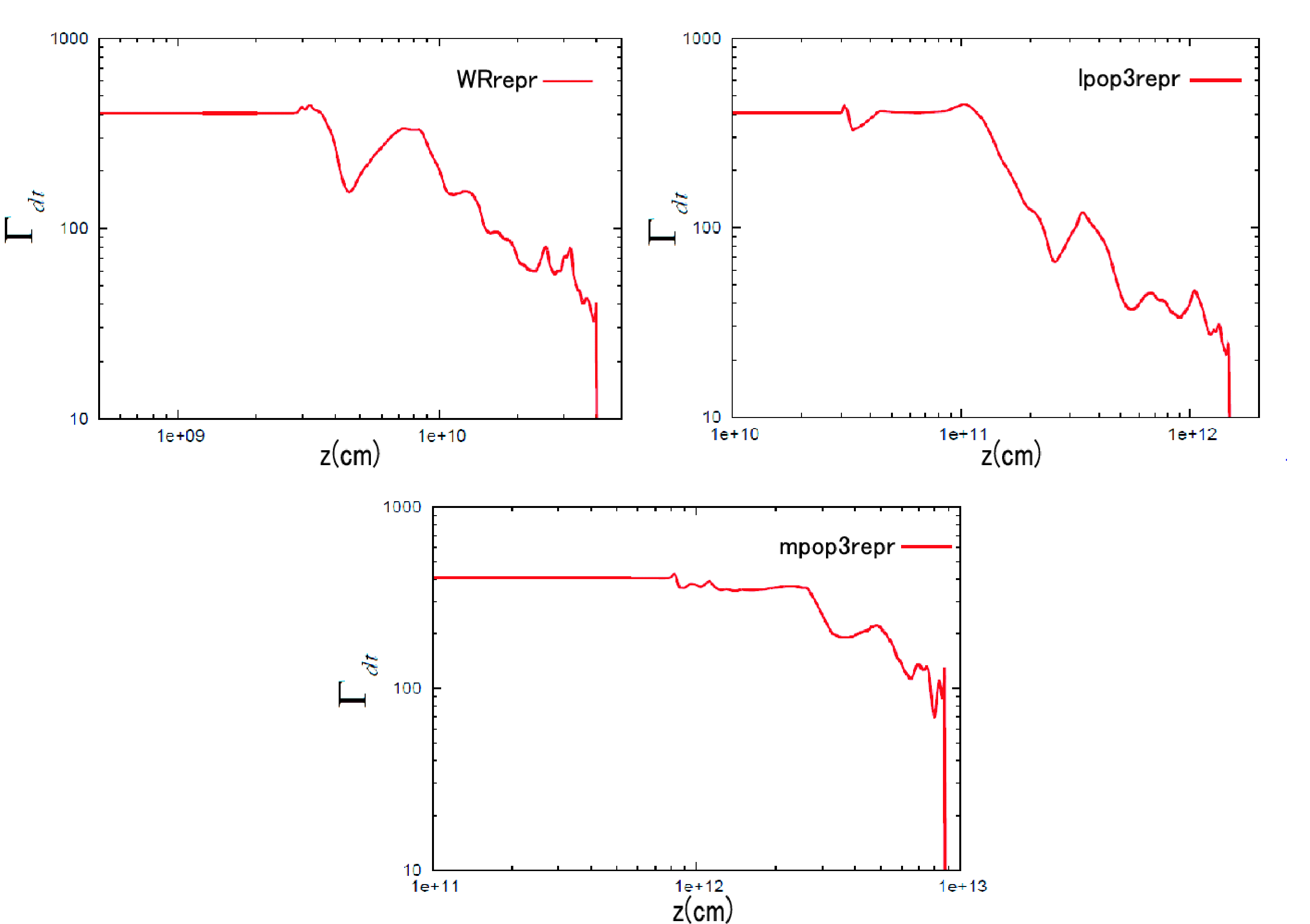}
\caption{Same as Figure~\ref{f7}, but for representative models.
\label{f15}}
\end{figure}

We summarize numerical results in Table~\ref{tab3}. As you can see in this table, the forward shock wave propagates faster than the reference model. We also find that both the total amount of injection energy and diagnostic energy at the breakout time are larger than those of reference models. Figure~\ref{f15} shows the $\Gamma_{dt}$ distribution along the jet axis at the shock breakout for each model. As you can see in this figure, every models produces more relativistic breakout than the corresponding reference models. In particular, it is interesting to note that mpop3repr succeeds highly relativistic breakout in spite of the large massive envelope. Note that the outer ejecta for WRrepr and lpop3repr is still mildly relativistic ($\Gamma_{dt} \sim 30 $). This is most likely caused by the numerical baryon contamination. However, even if the baryon pollution were real, these outflows could accelerate relativistically and create GRBs for the same reason discussed in Section~\ref{subsecdepeoneff} (see also Section~\ref{subseclimitation}).

% We also find that the reverse shock successfully move out (See Figure~\ref{ftempo4}}
% The jet power at break out for each model is $1.8 \times 10^{51}$erg/s for WRrepr, $6.6 \times 10^{49}$erg/s for lpop3repr and $8.6 \times 10^{49}$erg/s for mpop3repr, respectively.}
% Thus this fact supports the idea that the massive Pop III stars are progenitor candidates for GRBs.

% As you can clearly see in this figure, all models succeed to break out relativistically.

\subsection{Limitation of the current study} \label{subseclimitation}
%Although our simplification does not change the essence of our new findings, there are a lot of technical limitation for current numerical simulations. Here, we point out several limitations on the work presented in this paper.

In this subsection, we give some important cautions for the results presented in this paper. Although our simplification of numerical methods does not change the essence of our new findings, there are some technical limitations for the current numerical simulations. Here, we point out several limitations on the present work with some additional simulation results.
\subsubsection{Baryon pollution}
%The code which we are used in this paper was developed using the central-scheme.
Although our numerical code succeeds in capturing strong shock waves and complex turbulence in relativistic outflows, numerical diffusion is inevitably inherent in it. As a result, numerical diffusion potentially induces artificial baryon pollution which leads to a lower diagnostic terminal Lorentz factor in the jet. As we have already mentioned, since $\Gamma_{dt}$ is important in determining whether the jet would eventually create a strong relativistic outflow or not, we need to know how the numerical resolution affects of $\Gamma_{dt}$ along the jet axis.

\begin{figure}
\vspace{15mm}
\epsscale{1.0}
\plotone{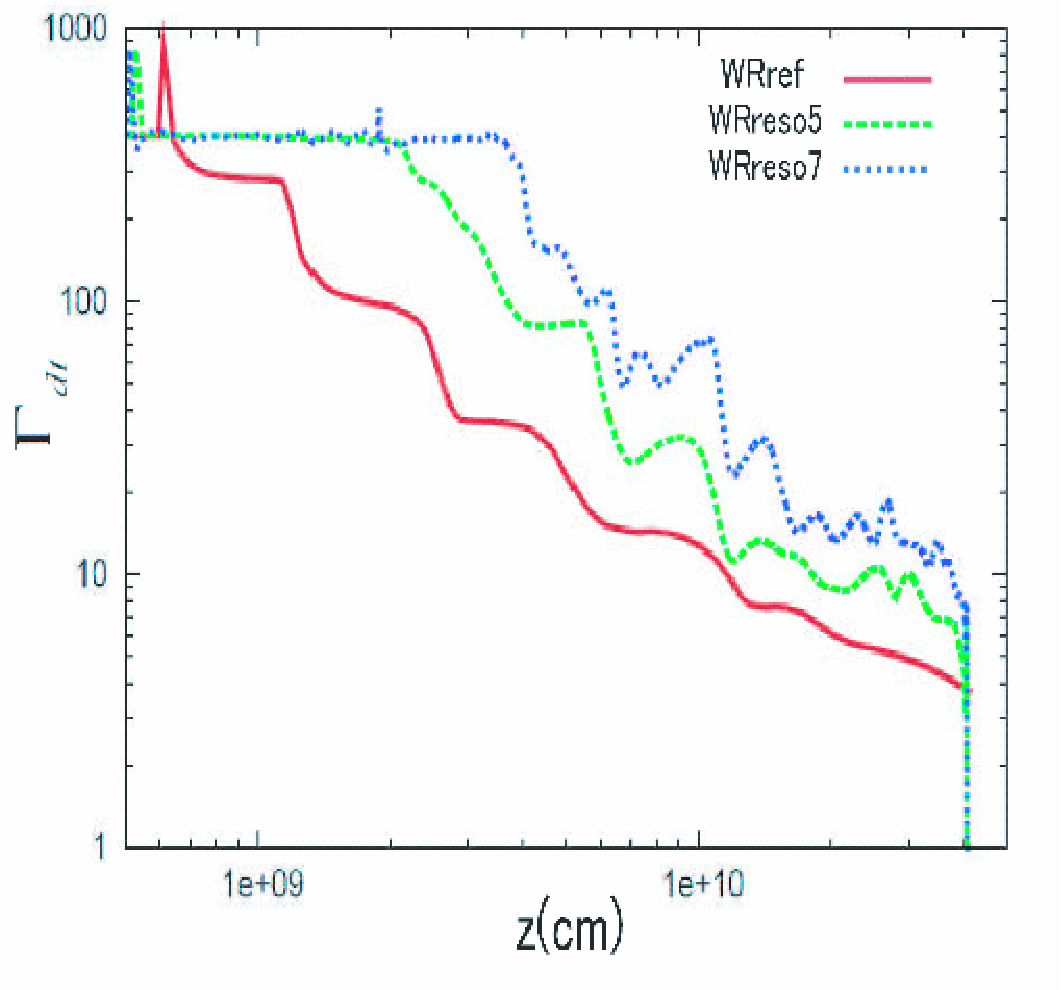}
\caption{Same as Figure~\ref{f7}, but with different spatial resolutions among WR models (WRref, WRreso5 and WRreso7).
\label{f16}}
\end{figure}

\begin{figure}
\vspace{15mm}
\epsscale{1.0}
\plotone{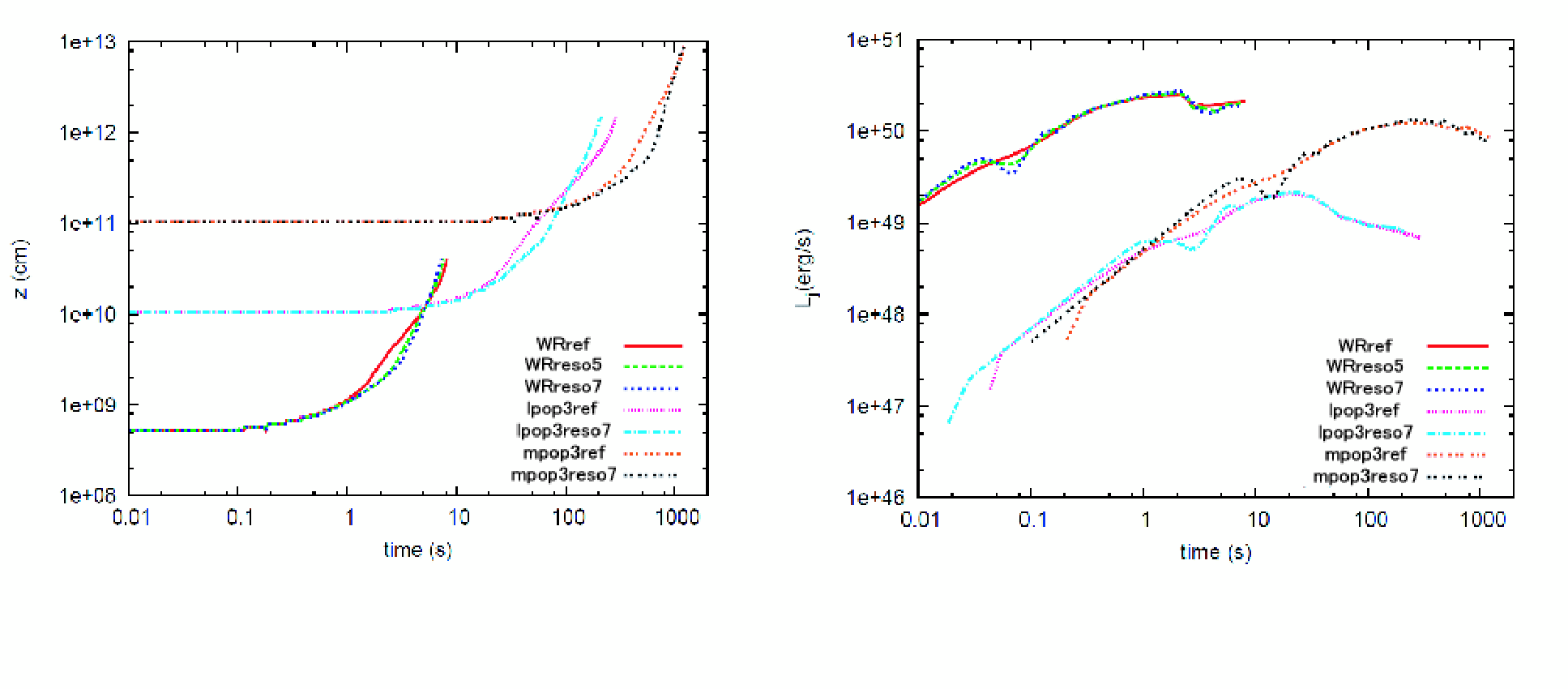}
\caption{The dependence on the resolutions of our simulations. Left:
  the forward shock evolution along the z-axis, Right; the time
  evolutions of the luminosity.
\label{f17}}
\end{figure}

 Figure~\ref{f16} shows the comparison of $\Gamma_{dt}$ distribution along the jet axis for models with different spatial resolutions. As you can see in this figure, models with higher resolution exhibit a larger $\Gamma_{dt}$. This confirms that higher resolution reduces the baryon pollution and shows the higher value of $\Gamma_{dt}$. In fact, the amount of baryon mass around the jet axis at the breakout time is decreasing with higher resolutions (see Table~\ref{tab4}). The baryon mass contained in the jet is roughly estimated as
\begin{equation}
M_{b} = 2 \pi (1 - \cos\theta_{op}) \int_{r_{in}}^{r_{out}} \rho_z (r) r^2 dr, \label{eq:baryonmass}
\end{equation}
where $\rho_z (r)$ denotes the density profile along the z-axis.
We also estimate the average terminal Lorentz factor $\Gamma_{f}$ for the jet as
\begin{equation}
\Gamma_{f} = \frac{E_{j}}{M_{b} c^2 }, \label{eq:averagetermgamma}
\end{equation}
where $E_{j}$ denotes the total injected energy after the breakout. We estimate $E_{j}$ as
\begin{equation}
E_{j} = \eta M_{bound} c^2 \label{eq:averagetermgamma}
\end{equation}
where $M_{bound}$ denotes the total mass of gravitationally bound matter (see also Eq.~\ref{eq;Elc} for the definition of bound matter). Here, we assume that all of energy from the central engine is successfully transferred to outgoing ejecta. As shown in Table~\ref{tab4}, the outflow for WRref can accelerate $\Gamma_{f} \sim 60$ although this model is heavily influenced by the baryon pollution.
% We also emphasize that the current estimation of the jet luminosity is affected by the location of the inner boundary and it would be larger than the current estimation (See below).
Thus, even if the baryon pollution were taking place in real, this model could create the relativistic outflow. In addition, we would like to emphasize that the macroscopic jet evolution is not sensitive to the numerical resolution (see Figure~\ref{f17}).

% Incidentally, Table~\ref{tab3} also shows that the later injection models (WRM3 and WRM6) tend to have higher the average terminal Lorentz factor than the reference model (WRref). Therefore, as discussed in Section~\ref{subinjeclorentzandtiming}, the later jet injection is more suitable for producing GRBs than the earlier one.
% \naga{We also emphasize} that the $\Gamma_{dt}$ distributions which are shown in Fig~\ref{f7}~and~\ref{f12} are the lower limits and the actual accretion-powered jet would break out more relativistically from the progenitor. It should be noted that the macroscopic jet evolutions are not sensitive to the spatial resolutions. As we can see in Figure~\ref{f17}, the time evolutions of the forward shock and luminosity are almost the same for different resolution models.

 We also point out that the two-dimensional axisymmetric setup affects baryon pollution. In our simulations, the baryon pollution near the pole is mainly attributed by the numerical diffusion, since the meridian velocity near the pole is almost zero due to the axisymmetric property. However, in reality, non-axisymmetric motions and finite meridian velocity may happen in the jet region. In this case, there is a possibility that the cocoon or strong back flows thrust into jets, then a lot of baryons could mix with the jet matter. Note also that the dense ``plug'' at the head of the jet is an artifact of the axisymmetric property. In reality, the non-axisymmetric motions of jet would cause dispersion of the dense ``plug'' \citep{2004ApJ...608..365Z}.

% Under the present studies, we can not obtain the precise amount of baryon in the jet due to above limitations. It should be noted that, \naga{even if our numerical simulations inherent artificial baryon pollutions, some models in this study could potentially create GRBs since the inner fast moving jet would catch up the forward slowly moving ejecta as mentioned in Section~\ref{subsecdepeoneff}.} Needless to say, if the baryon pollution is less efficient, outflows around polar regions can accelerate to the relativistic velocity and may create GRBs.
% In fact, the amount of baryon in the jet is not too much to create GRBs in our simulations (See in Table~\ref{tab3}).
% However, as mentioned in Section~\ref{subsecdepeoneff}, there is the possibility of GRB production even if the
% Based on \naga{the} above discussions, we can not give strong constraints whether the accretion powered jet would create mildly-relativistic \naga{explosions} or GRBs in the present study. In order to give strong constraints for them, we need to take into account further issues which are described below.
\subsubsection{Dependence on Inner Boundary Location }
One of the other drawbacks in the present study is the choice of inner boundaries in current models. For all simulations, the inner boundary is located well outside the central core of the star. As a result, the mass accretion rate for each model would be different from the actual accretion rate in the vicinity of black hole. In addition, the impact of feedback would also be changed. In order to remove these uncertainties, we demonstrate some numerical simulations for different position of the inner boundary.
%in order to understand the tendency of its influence.
% we check the dependence of the choice of inner boundary.

\begin{figure}
\vspace{15mm}
\epsscale{1.0}
\plotone{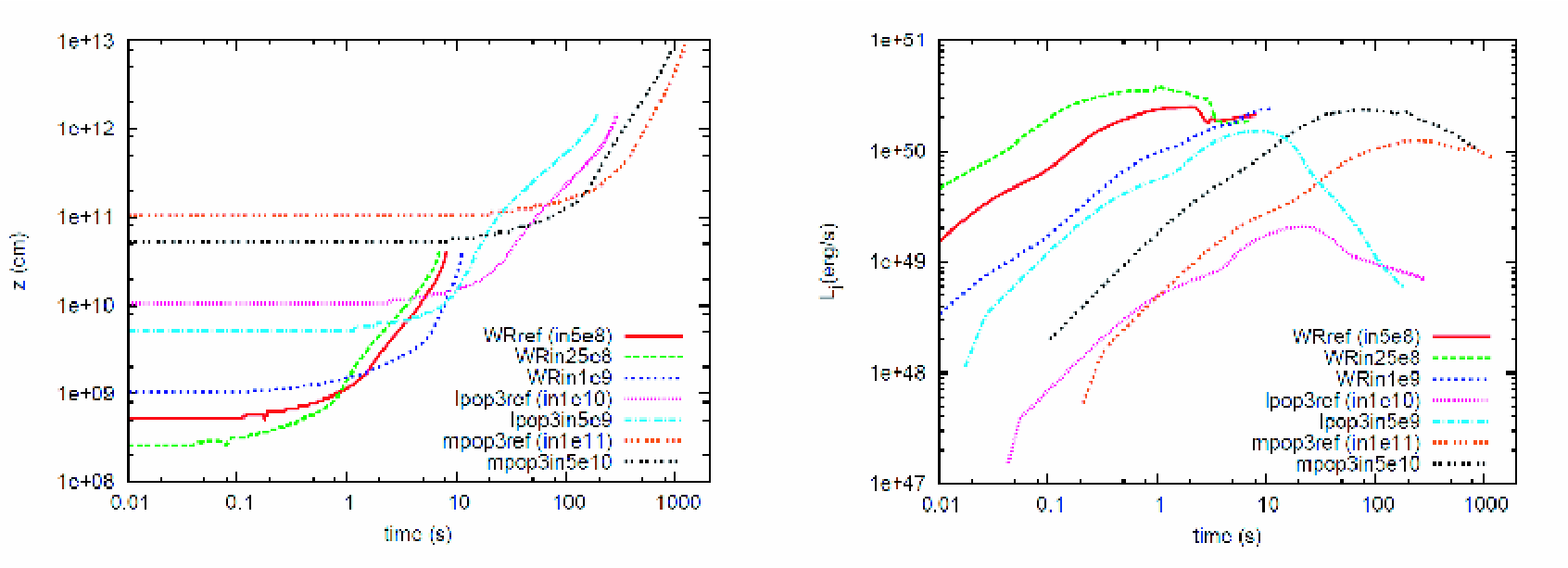}
\caption{The dependence on the location of the inner boundary. Left:
  the forward shock evolution along the z-axis, Right; the time
  evolutions of luminosity.
\label{f18}}
\end{figure}

Figure~\ref{f18} shows the dependence on the inner
boundaries for the evolution of the forward shock wave on the z-axis
and the jet luminosities.
As shown in these panels, models with a smaller inner boundary imply that the forward shock wave propagates faster. Since the density is higher at the inner radius, the accretion rate consequently becomes larger so that the jet luminosity also becomes larger. In these panels, we find that the model lpop3in5e9 has a peak accretion rate of about 10 times that of model lpop3ref, while WR and mpop3 simulations exhibit only the factor of 2 or 3 amplification, when the inner boundary is smaller by a factor of 2.
% while the WR and mpop3 simulations only lead to an increase of 2 or 3 in the accretion rate when their inner boundary is moved in by a factor of 2.
% It indicates that the inner boundary is particularly important for lpop3 models.
 We also note that, for the mpop3 models, the initial enclosed mass at the inner boundary is 414 $M_{\sun}$ and it corresponds to nearly half the total mass of the star. Thus, the location of the inner boundary substantially affects the jet dynamics especially for lpop3 and mpop3. In order to obtain more quantitative arguments for the actual jet luminosity and breakout time, we need the simulations with a smaller inner boundary, which are beyond the scope of this paper.
% Although we confirm that the location of the inner boundary strongly affects the jet dynamics, the effect is quite systematic:

It should be noted, however, that the effect of location of the inner boundary is
 quite systematic:
 the forward shock wave becomes fast and the
luminosity becomes large if we put the inner boundary on a small radius.
This is probably because the density is initially higher near the boundary.
Consequently the total mass accretion rate increases and therefore the jet luminosity becomes large with a strong forward shock wave.
According to these results, it is a robust claim that the shock breakout
is possible if the accretion-to-jet conversion efficiency satisfies the condition $\eta \gtrsim 10^{-4}$ (see more details in the subsection
\ref{sec:ana}).
%It should be noted, however, that the model lpop3in5e9 has a peak accretion rate of about 10 times that of model lpop3ref while the WR and mpop3 simulations only lead to an increase of 2 or 3 in the accretion rate when their inner boundary is moved in by a factor of 2. It indicates that the inner boundary is particularly important for lpop3 models. In addition, for the mpop3 models, the initial enclosed mass at the inner boundary is 414 solar masses and it corresponds to nearly half the total mass of star. 
\subsubsection{The time delay between infall and injection}
In the present study, we assume that the jet outflow is injected immediately after the mass inflows through the inner boundary. However, it takes a certain amount of time for the fluid to reach the vicinity of the black hole in reality. In addition, the outflow should travel up to the inner boundary. The time difference can be roughly estimated based on the assumption that the matter is free fall to a black hole. At the beginning of our simulation, it takes $\sim 1~s$ for WR model from the inner boundary to the center of the core, while $\sim 20~s$ and $\sim 150~s$ for lpop3 and mpop3, respectively.  Our immediate energy injection would be different from reality and potentially affect our findings. Here, we investigate the influence of the time delay between infall and jet injection.

We perform numerical simulations for some models with taking into account the delay time of the jet injection. In this study, we estimate the delay time as free-fall time from the location of the inner boundary to the Schwartzschild radius. Note that the Schwartzshild radius is estimated by the enclosed mass at the inner boundary. We ignore the jet propagation time from the vicinity of the black hole to the location of the inner boundary. This is because the time scale of jet propagation is much less than the free-fall time. The jet injection parameters, such as the conversion efficiency and opening angles, are the same as the reference models. Hereafter, we denote these models as WRdelay, lpop3delay and mpop3delay for WR, lpop3 and mpop3 progenitors, respectively.

In Table~\ref{tab3}, we show the summary of results for these models. In comparison with reference models, the overall dynamics are qualitatively similar to those of reference models. In addition, the variation of the breakout time, the total injection energy, and the diagnostic energy remain within $10\%$ from those of reference models. Thus we confirm that the time delay between infall and injection is a minor effect. This seems to be attributed to the fact that the accretion rate is almost constant except for at the very early phase of collapse. As a result, the delayed time injected jet goes through comparable ram pressure to the non-delayed injection case. Therefore the forward shock evolution is similar to the reference models.
% Based on above arguments, we need to take into account the effects of the location of inner boundary rather than the delay time. Needless to say, if the location of inner boundary becomes the vicinity of central engine, we do not have to care about the delay time.
% Based on above discussions, we need to care about the location of inner boundary rather than the time delay for more quantative arguments.
%we would take into account the position of inner boundary rather than the time delay. Needless to say, % In addition, overall dynamics do not change so much in these simulations. It should be noted that the effect of time delay is also due to the position of our inner boundary. If we can put them near the black hole, the effect of time delay are also reduced. Thus, for more quantative arguments, we need to perform simulation with smaller inner boundary which is our future work.
\subsubsection{Long term simulations}
% In addition, it is still unclear whether the pure jet component can \naga{progress anteriorly} after the shock breakout. The contact discontinuity between pure jet matter and stellar mantle should progress anteriorly.
% Thus, for the complete understanding of these issues, the long term numerical simulations are strongly required.
In the current study, we investigate the jet propagations inside of the stellar mantle and discuss the possibility of GRBs. However, in order to know how these outflows accelerate in the ISM, it is necessary to carry out simulations with longer duration and a large spatial region. Although our simulations have a limited duration and spatial extent, we perform extended numerical simulations for the WR reference model for the purpose of qualitative understanding of the outflow properties. We extend the computational boundary to $r \sim 5 R_{star}$ and carry out the simulation until the forward shock wave reaches the outer boundary. We cover 1500 uniform radial grids in this extended spatial region, i.e. the total number of our radial grids are 2000 (500 + 1500). The grid width is the same as the outermost radial width of previous calculations. The outer boundary in this calculation is located at $r_{out} = 2.34 \times 10^{11} {\rm cm}$. In this simulation, the 3-level AMR technique is employed. Thus the radial resolution in the extended region corresponds to $\Delta r = 4.27 \times 10^{7} {\rm cm}$. The meridian grid width is the same as previous calculations.

\begin{figure}
\vspace{15mm}
\epsscale{1.0}
\plotone{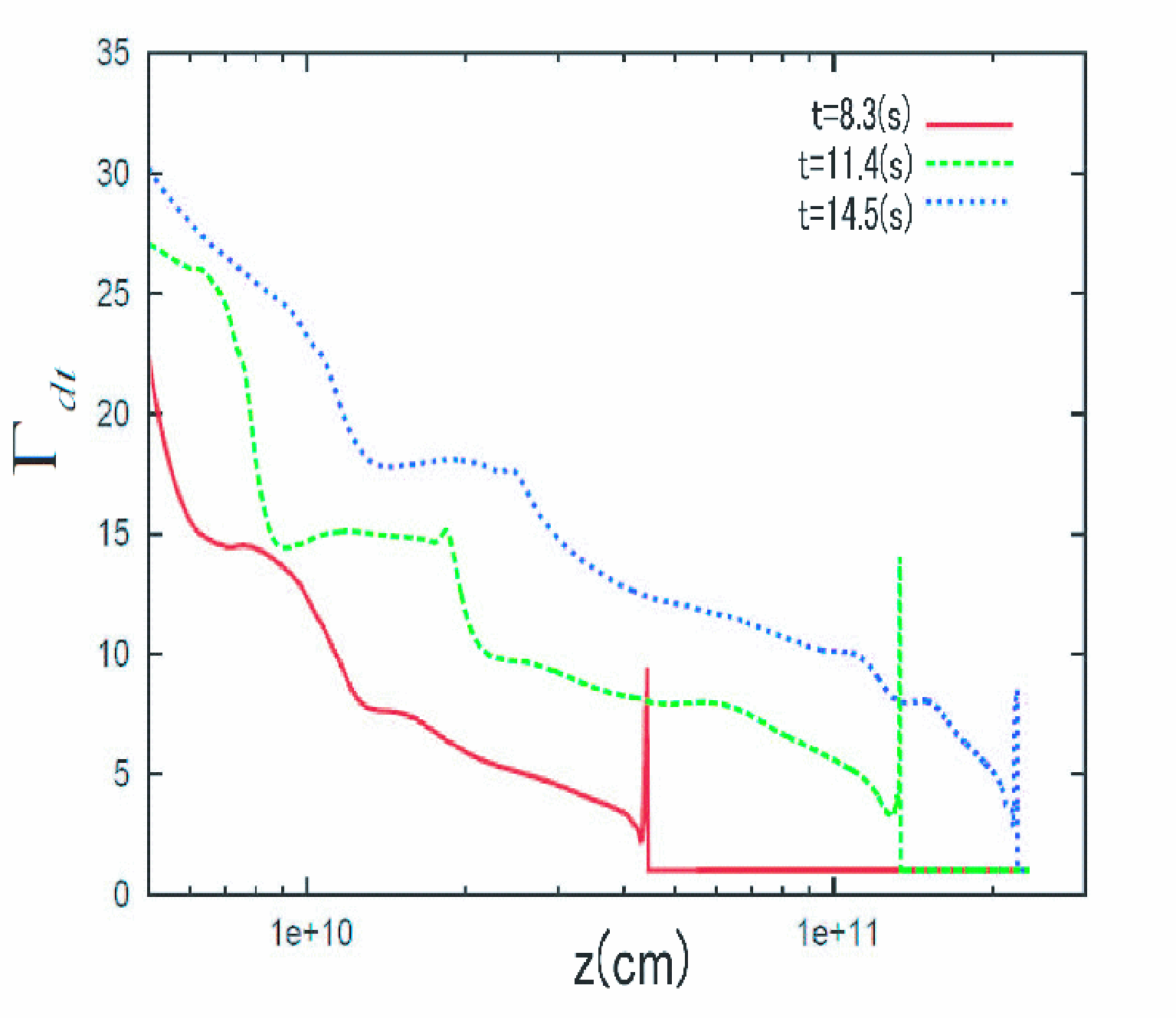}
\caption{Same as Figure~\ref{f7}, but long term simulations for WRref. The red line indicates the diagnostic terminal Lorentz factor distribution soon after the shock breakout, while the blue line indicate the same distribution but $t=14.5(s)$ which is the final time of this simulation. The snapshot for green line ($t=11.4(s)$) corresponds to the middle time between them.
\label{f19}}
\end{figure}

 Figure~\ref{f19} shows some snapshots for diagnostic terminal Lorentz factor distribution along jet axis. As we can see in this figure, the outflow propagates into this ISM and the inner ejecta with high $\Gamma_{dt}$ gradually progresses outward with time. This is attributed to the continuous energy injection of the central engine. According to these results, even if the baryon pollution were the real, the outflows would continuously propagate into the ISM because it is being pushed by high Lorentz factor ejecta in the back. Note also that, since this simulation is contaminated by the artificial baryon pollution, the outflow is more relativistic in reality than this simulation. The encouraging tendencies increase the possibility of GRBs being produced in a wider range of scenarios, but future lon-duration simulations are needed.

In summary of this subsection, in order to judge whether the outflows become relativistic or not in reality, we need to simulate them with (1) high resolution in order not to affect the baryon pollution, (2) smaller inner boundary, and (3) long duration and large spatial range. Here, we emphasize that the results of current study give conservative claims for the possibility of relativistic jet breakout, which is the minimum requirement for producing GRBs. In fact, we show that the accretion-powered jet succeeds to break out relativistically from massive Pop III progenitors if the adequate conditions are satisfied.
\subsection{Comparison with Analytic Estimate}
\label{sec:ana}
In this subsection, we compare our numerical results with the analytic
estimate of the shock breakout
%through the massive envelope obtained 
in \cite{2011ApJ...726..107S} and Appendix \ref{sec:appendix_b}.
%which investigated the
%  penetrability of the jet through the massive envelope using analytic way.
Since setups are rather different between the numerical study and 
the analytic one, we concentrate on the parameter dependence of the successful relativistic shock breakout.
They derived the shock breakout time ($t_\mathrm{b}$; Eq. 12
in their paper) and the free-fall timescale of the envelope materials,
i.e., the duration of the accretion-powered jet ($t_\mathrm{ff}$;
Eq. 15).  When $t_\mathrm{b}$ is shorter than $t_\mathrm{ff}$, the jet
successfully arrives at the stellar surface and the relativistic shock
%\naga{??? jet or shock ???}
breakout takes place.  Employing 
%their Eqs. (12) and (15), 
Eqs. (\ref{eq:tfftb}) and (\ref{eq:lambda}) in Appendix \ref{sec:appendix_b},
we can
derive the criteria for the successful relativistic shock breakout as
\begin{equation}
%\frac{t_\mathrm{ff}}{t_\mathrm{b}}\sim \alpha\left(\frac{\eta}{10^{-3}}\right)^{0.79}\left(\frac{\theta_{op}}{5^\circ}\right)^{-1.6}\gtrsim 1,
%\frac{t_\mathrm{ff}}{t_\mathrm{b}}\sim \alpha\left(\frac{\eta}{10^{-3}}\right)^{3/(9-2n)}\left(\frac{\theta_{op}}{20^\circ}\right)^{-6/(9-2n)}\gtrsim 1,
\frac{t_\mathrm{ff}}{t_\mathrm{b}}\sim \lambda\left(\frac{\eta}{10^{-4}}\right)^{\frac{3}{9-2n}}\left(\frac{\theta_{op}}{5^\circ}\right)^{-\frac{6}{9-2n}}\gtrsim 1,
\label{eq:tff_tb}
\end{equation}
%\naga{??? in the abstract, we use $\eta\sim 10^{-4}$ and $\theta_{op}
 % \sim 20^{\circ}$. why don't you use these values here ???}  
where
$n$ ($\approx 2.6$ for mpop3 and WR and $\approx 2.1$ for lpop3) is an
index for the density profile of the stellar envelope,
\begin{equation}
\rho(r) \propto \left(\frac{R_*}{r}-1\right)^n,
\label{eq:rho}
\end{equation}
with the stellar radius $R_*$ \citep{1999ApJ...510..379M}.  The
prefactor $\lambda$ depends on the stellar radius, the core mass and
the envelope mass,
%\naga{in Eq.~(\ref{eq:lambda}),}
%(\naga{see \citealt{2011ApJ...726..107S} and Appendix \ref{sec:appendix_b}}), 
and is found to
%be of order unity, 
be typically around 
%$\lambda \sim 0.5$
$\lambda \sim 3$
 for mpop3
according to \cite{2011ApJ...726..107S} 
and Appendix \ref{sec:appendix_b}.
In this paper, we find that
%$\lambda \approx 1$ 
$\lambda \approx 2$
is suitable for explaining the numerical results of
mpop3.
The reason of the difference in $\lambda$ is that some of
the assumptions in \cite{2011ApJ...726..107S} are violated, that is, they
assumed a conical jet propagation (i.e., $\theta_{op}$ is independent
of the radius; see \citealt{2011PThPh.126..555I}) and neglected the
negative feedback from the cocoon on the accretion rate.
The breakout condition is also not exactly the same
between the analytic and numerical calculations.
Nevertheless it is encouraging that
both results coincide within a numerical factor.

%the order of magnitude.
%%% added by YS 12/3
%%% deleted by YS 14.2.2012
%It should be noted that it can be confirmed by this scheme that a red
%supergiant (RSG) is not a candidate of a GRB. We can roughly estimate
%$\lambda$ for a RSG and obtain $\lambda\sim 5$\footnote{In this
%  estimate, we employ the core mass of $M_c=5M_\odot$, the envelope
%  mass of $M_\mathrm{env}=10M_\odot$, and the core radius of
%  $R_c=3\times10^{10}$ cm, respectively. It should be noted that the
%  density index $n$ of a RSG is different from mpop3, but the order of
%  $\lambda$ would stay in the same order.} that is much larger than
%mpop3. Therefore, $\lambda$ would be a good indicator for the
%possibility of GRB production.

\begin{figure}
\vspace{15mm}
%\epsscale{0.5}
\epsscale{0.4}
%\plotone{f16_1.eps}
%\plotone{f16_2.eps}
%\plotone{f16_3.eps}

\plotone{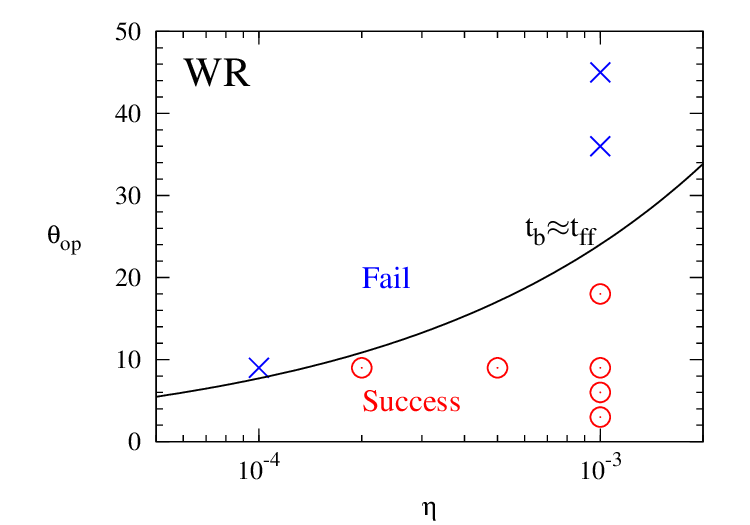}
\plotone{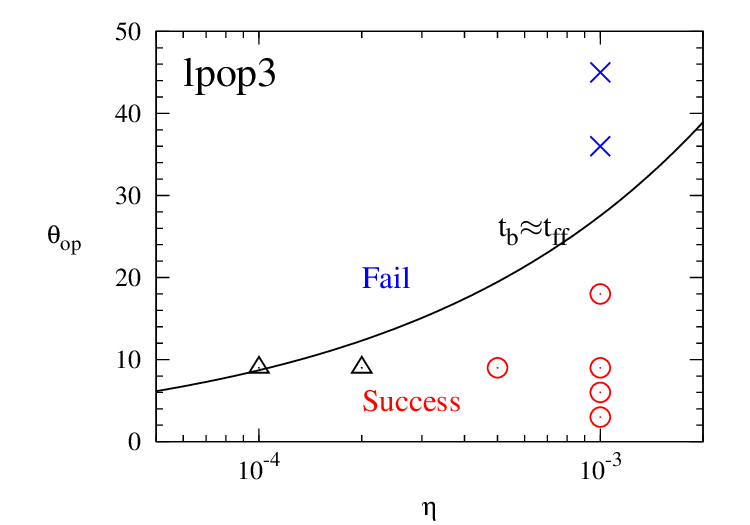}
\plotone{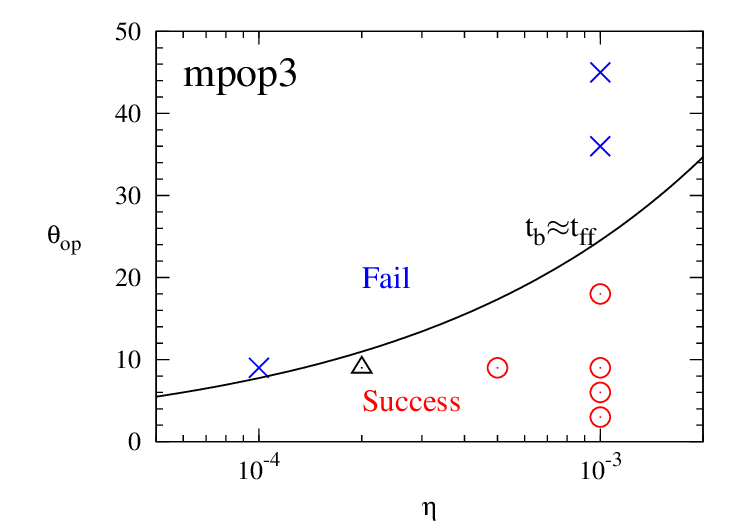}

\caption{The score sheet of the shock breakout for WR (top panel), lpop3
  (middle panel), and mpop3 (bottom panel), respectively,
  in the accretion-to-jet conversion efficiency $\eta$
  and the opening angle $\theta_{op}$ plane. Red circles
  correspond to the models which have possibilities for creating GRBs
 in our numerical
  simulations, while blue crosses show the failed cases.  
  Black triangles are marginal models (see text for details).
  The
  analytical criteria for the shock breakout in Eq. (\ref{eq:tff_tb})
  are shown by the black solid lines.  Below this line the shock wave can
  break out of the stellar surface before the mass accretion
  ceases. On the other hand, the jet stalls inside the massive
  envelope and the explosion becomes spherical above this analytical
  line.
\label{f20}} 
\end{figure}

In Figure \ref{f20} we present $\eta-\theta_{op}$ diagrams to compare
our numerical results with the analytic criteria in
Eq.~(\ref{eq:tff_tb}) (black solid line). The circles, triangles and
crosses in this figure correspond to the column of
``\textit{Possibility of GRB production}'' in Table~\ref{tab3}.
%In Figure \ref{f20} we present $\eta-\theta_{op}$ diagrams to
%compare our numerical results of the successful breakout (red circles)
%and the failed explosion (blue crosses) with the analytic criteria
%in Eq. (\ref{eq:tff_tb}) (black solid line).
We use $\lambda=2$ for all progenitor models.
 As we can see from
Figure \ref{f20}, the analytical criteria in Eq. (\ref{eq:tff_tb}) and
\cite{2011ApJ...726..107S} are useful to forecast the penetrability of
the relativistic jet and the possible GRB production in massive star
models. We can predict the parameter space where GRBs could occur in
$\eta-\theta_{op}$ plane once we calibrate $\lambda$ with numerical
simulations.  
In addition, we find that the analytical critical curve can be approximated by a much simpler form for all current progenitors (WR, mpop3, lpop3) as
%the simple analytical criteria of the
%  possibility of GRBs for all stellar models in this paper when the
%  accretion-to-jet conversion efficiency satisfies
%We find the following relation, which is much simpler, irrespective of the progenitor as
\begin{equation}
\eta \gtrsim 10^{-4} \left(\frac{\theta_{op}}{8^{\circ}}\right)^2.
\label{eq:criteria}
\end{equation}
%\naga{??? how do we derive 8 degree? do we use lambda 2.6? ???}

Note that a red supergiant (RSG) 
satisfies the breakout criteria in Eq.~(\ref{eq:tff_tb}) with $\lambda \sim 6.8$.
Actually a GRB jet might be associated with a RSG,
which has not been observed so far.
A GRB from a RSG would be very dim and long
because the breakout time is long due to the expanded envelope
and the accretion rate is low at the late time.
Such dim GRBs might even dominate as the sensitivity is improved.
Alternatively, another condition could prevent the GRB production.
First, the velocity of the jet head should be faster than 
that of the cocoon, $v_h>v_c$ \citep{2003MNRAS.345..575M,2011ApJ...726..107S}.
In some parameter space, a RSG does not satisfy this second criterion
because of its shallow envelope
and the outflow becomes almost spherical
\citep{2011ApJ...726..107S}.
Second, the central engine of a RSG might not rotate
fast enough for the jet production.
The detailed discussion will be presented in the forthcoming paper.

\section{Conclusions}
In this paper, we investigate the propagation of the
accretion-powered jets in Pop III and present-day stellar
progenitors (Wolf-Rayet stars). 
We perform two-dimensional relativistic hydrodynamic
simulations taking into account both the envelope collapse and the jet
propagations for the first time.  The main findings in this paper are
summarized as follows.

\begin{enumerate}
\item Although some portion of matter ceases to fall and is pushed outward by
  the energy injection from the polar region, a large amount of matter
  continues accreting into the black hole, so that the central engine can
  continue to inject outflows.

\item If the central engine satisfies a certain condition (see
  below), the jet can successfully propagate and create a relativistic
 breakout from various types of progenitors for GRBs. In particular, we show that Pop III stars could be progenitor candidates for GRBs, as pointed out by \citet{2011ApJ...726..107S}.
% especially from the massive ($\sim 1000 M_{\odot}$) Pop III stars, as pointed out by \citet{2011ApJ...726..107S}.
  We numerically verify that the central
  engine can last very long $ \gtrsim 100$ s for light Pop III stars and  $ \gtrsim 1000$ s for massive Pop III stars because of the accretion of the huge envelope.

\item
% The outflow succeeds the jet breakout
The jet can produce a relativistic breakout if the accretion-to-jet
conversion efficiency satisfies 
\begin{equation}
\eta \gtrsim 10^{-4} \left(\frac{\theta_{op}}{8^{\circ}}\right)^2,
\label{eq:criteria2}
\end{equation}
as derived in Eqs.~(\ref{eq:tff_tb}) and (\ref{eq:criteria}).
%% \begin{equation}
%% \eta \gtrsim 10^{-3} \left(\frac{\theta_{op}}{20^{\circ}}\right)^2,
%% \label{eq:criteria}
%% \end{equation}
%as derived in Eq.~(\ref{eq:tff_tb}).
%If the accretion-to-jet conversion efficiency satisfies the
%  condition $\eta \gtrsim 10^{-4}$, .
  % Otherwise the injection energy advected inward with infalling matter.
 Otherwise the injection energy results in non-relativistic
% spherical
 explosion or advection into a black hole.
%We derive Eq.~(\ref{eq:tff_tb}) with \citet{2011ApJ...726..107S}, and compare
%  it with our numerical results in Figure \ref{f20} on the plane
%  of the accretion-to-jet conversion efficiency $\eta$ and the half
%  opening angle of jets $\theta_{op}$. We find that the analytical
%  criteria for the jet breakout are consistent with our numerical
%  results.

%% \item We also investigate the dependence on the opening angle of the
%%   outflows, and summarize the breakout criteria 
%%   in Eq.~(\ref{eq:criteria}).
%% %that the jet potentially creates GRBs if
%% % $\theta_{op} \lesssim 20^{\circ}$.
%%   On the other hand, for very wide
%%   jets ($\theta_{op} = 36^{\circ}, 45^{\circ}$), these dynamics are
%%   very complicated and sometimes the catastrophic fallback makes
%%   strong outflows from the central engine.  However, we find that they
%%   might not be candidates for GRBs since the diagnostic terminal
%%   Lorentz factor is too low to create GRBs.

\item We find that the timing of jet injection does not affect our results significantly. We also observe that the diagnostic terminal Lorentz factor
  tends to be higher for later jets than for early jet injection
  because the rarefaction wave reduces the central density before the strong jet is injected.
% Thus, we confirm that, even if the operation of central engine is somewhat late, the accretion-powered jet accomplishes the shock breakout and potentially creates GRBs.
%  Thus we can
%  conclude that the late time operation of central engine is more suitable for producing
% GRBs than the early one.

%\item We suggest that the delayed jet injection is preferred for
%  producing GRBs.
% At least for Wolf Rayet progenitors, the best
%  condition seems that the central engine starts to operate at the
%  time when the black hole mass becomes $3 M_{\sun} \lesssim M
%  \lesssim 6 M_{\sun}$.

% Although we confirm that the location of inner boundary affects our
% findings quantatively, our findings in this paper do not change.

%\item We derive the analytical criteria for the jet breakout in
%  Eq.~(\ref{eq:tff_tb}) with \citet{2011ApJ...726..107S}, and compare
%  it with our numerical results in Figure \ref{f20} on the plane
%  of the accretion-to-jet conversion efficiency $\eta$ and the half
%  opening angle of jets $\theta_{op}$.  We find that the analytical
%  criteria for the jet breakout are consistent with our numerical
%  results.
\end{enumerate}

It should be noted that our simulations are affected by numerical diffusion and the location of inner boundary as analyzed in Section~\ref{subseclimitation}.  Especially for Pop III models, we put the inner boundary far outside the central core. We intend to address these issues in future publication.

We would like to point out that the injected energy from the central
engine may contribute to the explosion energy of the supernova
component, if the hot cocoon mainly contributes to the supernova,
i.e., an almost spherical shock wave accompanying a large amount of 
nickel production (See also \citet{2002MNRAS.337.1349R} for discussions of excess energy accumulated in the cocoon).
If the strong reverse shock wave is still stagnated
around the root of the jet at the time of the shock breakout, the
kinetic energy of the jet continues to be converted into thermal
energy deep inside of the stellar mantle.
% ??? please clarify this sentense. it is unclear for me ??? what is
% the ternminal shock ???}.
Although the cocoon pressure would decrease after the shock breakout
and make it easy for the reverse shock wave to propagate, the jet
luminosity is also reduced as the accretion rate decreases, so that
the reverse shock wave may stay deep inside the star. If this is the
case, the injected energy may work to expel the stellar mantle
rather than contribute to the GRB component.  However, at the
present time, it is unclear how large an amount of nickel is created in
the hot cocoon, so we do not know whether the cocoon creates the supernova or
not. We will address these issues in a forthcoming paper.

It is also interesting to note that
%We also find that
if the stellar envelope consists of multi-layerd configurations 
(i.e. an onion like structure), the accretion rate also changes across the
different layer.
%% In our simulations, the rapid decline of the
%% accretion rate can be seen due to the discontinuous decline of the
%% density at the boundary of the shell (see e.g. Fig~\ref{f5}).
The change of the accretion rate may cause time variability of the central engine. The variability timescale depends on the initial location of the shell
boundaries and roughly corresponds to the free-fall timescale at this
region. It should be noted that the stellar rotation may play an
important role on this matter for the realistic situation since the
specific angular momentum is also discontinuous at the shell interface,
 and it may cause the accretion rate to fluctuate. 
The angular momentum is
generally an increasing function of radius in each shell, but it
discontinuously drops at the shell boundary. As a result, the
centrifugal force is reduced in strength there and it may potentially
increase the accretion rate.
% The study of whether this process happens
%or not is the beyond the scope of this paper. We will investigate
%these issues in the forthcoming paper.

It is interesting to note that some models in the present study may explode as choked GRBs. Although they can not produce GRBs, they are expected to produce TeV neutrinos \citep{2001PhRvL..87q1102M}. These observational signals provide important information on jet propagation in the stellar mantle. We would like to investigate the observational consequence of the different jet injection parameters in the forthcoming paper.

Finally, we would like to emphasize that
 the rotation profile of the stellar progenitor
 also affects the GRB production.
This is because the angular momentum profile may determine
not only the starting time of the central engine
 but also the disk conditions \citep{2006ApJ...641..961L,2009MNRAS.398.2005Z,2009ApJ...692..804L}.
The dependence on angular momentum profile of progenitor is currently being investigated \citep{2012nagetal}.
Additionally, there is no guarantee
that the jet luminosity is proportional to the accretion rate as assumed in
this paper.  Actually, if the neutrino mechanism plays a key role for
producing the jet, the jet luminosity also depends on the mass of the black
hole. In addition, the relation between the mass accretion rate and jet luminosity in neutrino mechanism would be different from those used in the present paper \citep[see e.g.][]{2011MNRAS.410.2302Z}. These areas present the opportunity for future inquiries and are currently being investigated.

%In addition, the dependency on the mass accretion rate would be different \citep[see e.g.][]{2011MNRAS.410.2302Z}.

% Note that the terminal Lorentz factor among our models is set as
% $\Gamma_{term} \sim 400$.

\acknowledgments
We are grateful to the anonymous referee for beneficial comments. We would also like to thank Mr. Rhosuke Hirai for useful comments and proofreadings. This work is supported in part by Grant-in-Aid for the Scientific Research from the Ministry of Education, Culture, Sports, Science and Technology (MEXT), Japan [Nos. 24740165,21684014,19047004, \\
22244019,22244030,23840023], and by HPCI Strategic Program of Japanese MEXT, and the Center for the Promotion of Integrated Sciences (CPIS) of Sokendai.

% This work is partially supported by the Japan Society for Promotion
% of Science (JSPS) Research Fellowships, Grant-in-Aid for the
% Scientific Research from the Ministry of Education, Culture, Sports,
% Science and Technology, Japan [Nos. 222913, 22740178, 21540281,
% 19104006]. This study is also supported by Program for Improvement
% of Research Environment for Young Researchers from Special
% Coordination Funds for Promoting Science and Technology (SCF)
% commissioned by the Ministry of Education, Culture, Sports, Science
% and Technology (MEXT) of Japan.

\appendix
\section{Appendix}
\subsection{Analytic stellar model}
%\section{Analytic stellar model}
\label{sec:appendix_a}

Using Equation (9) of \cite{2011ApJ...726..107S}, we can calculate the stellar radius $R_*$ as
%\naga{??? The order of parentheses is changed. ???}
\begin{equation}
R_*=10^{13}~\mathrm{cm}
\left(\frac{10^{\frac{1.2}{3-n}}}{10^{3}}\right)
\left(\frac{0.4}{3-n}\right)^{-\frac{1}{3-n}}
\left(\frac{R_c}{10^{10}~\mathrm{cm}}\right)
\left(\frac{M_c}{400 M_\odot}\right)^{-\frac{1}{3-n}}
\left(\frac{M_\mathrm{env}}{500 M_\odot}\right)^{\frac{1}{3-n}},
\label{eq:R*}
\end{equation}
where we calibrate the overall factor to reproduce the mpop3 radius
with $n=2.6$. This equation is a general form of Equation (10) of \cite{2011ApJ...726..107S}. 
From Eq.~(\ref{eq:R*}), we can estimate the stellar radii of WR, lpop3, and RSG as $8\times 10^{10}$ cm, $7\times 10^{11}$ cm, and $2\times 10^{13}$ cm, respectively, which well reproduce actual values
%, remaining the difference 
within a factor of 2-3. In these estimations, we employ following parameters; 
\begin{itemize}
\item WR: $n=2.6$, $M_c=11 M_\odot$, $M_\mathrm{env}=2 M_\odot$, and $R_c=10^{10}$ cm.
\item lpop3: $n=2.1$, $M_c=15 M_\odot$, $M_\mathrm{env}=25 M_\odot$, and $R_c=10^{10}$ cm.
\item RSG: $n=1.5$, $M_c=4 M_\odot$, $M_\mathrm{env}=8 M_\odot$, and $R_c=10^{12}$ cm.
\end{itemize}

%\section{Stellar dependence of breakout criteria}
\subsection{Stellar dependence of breakout criteria}
\label{sec:appendix_b}

In this section, we derive analytic expression for the breakout criteria. Note that we 
employ 
%several typical values 
typical sets of parameters
to normalize physical quantities (e.g., $\eta$, $\theta_{op}$). 
%These values are not selected to represent typical progenitor, but to vanish the dependence of the overall factors on the density slope parameter, $n$. 
In order to 
%estimate the quantities for the considered progenitor, 
apply to the individual progenitor,
one should insert the values in Appendix \ref{sec:appendix_a}.

Here, we approximate the density profile of the envelope as
\begin{equation}
\rho(r)=\rho_1\left(\frac{R_*}{r}\right)^n,
\end{equation}
in which the term of unity in Eq. (\ref{eq:rho}) is neglected for simplicity. This approximation is valid for $r\ll R_*$.
The free-fall timescale of a mass shell at the radius $r$ and the enclosed mass $M_r$ is given by
\begin{equation}
t_{ff}=\sqrt{\frac{r^3}{GM_r}},
\label{eq:t_ff}
\end{equation}
where $G$ is the gravitational constant.
% and $M_r$ is the enclosed mass within $r$.
The mass accretion rate is given by
\begin{eqnarray}
\dot M=\frac{dM_r/dr}{dt_{ff}/dr}\approx \frac{8\pi}{3}\frac{r^3}{t_{ff}}\rho_0\left(\frac{R_0}{r}\right)^{n},
\label{eq:mdot}
\end{eqnarray}
where $R_0$ is the arbitrary characteristic radius and $\rho_0=\rho_1(R_*/R_0)^n$.
Here we neglect the derivative of $M_r$ with respect to $r$ for $dt_{ff}/dr$.
Using Eq. (\ref{eq:mdot}), the jet luminosity emitted from the central object is written as
\begin{eqnarray}
L_j=\eta\dot M c^2=
%7.5\times 10^{49}~\mathrm{erg~s^{-1}}
7.5\times 10^{45}~\mathrm{erg~s^{-1}}
\left(\frac{\eta}{10^{-4}}\right)
\left(\frac{t_{ff}}{1~\mathrm{s}}\right)^{-1}
\left(\frac{r}{10^{11}~\mathrm{cm}}\right)^{3}
\left(\frac{R_0}{r}\right)^{n}
\left(\frac{\rho_0}{10^{-5}~\mathrm{g~cm^{-3}}}\right),
\label{eq:L_j}
\end{eqnarray}
where $t_{ff}$ corresponds to the jet breakout time at $R_0$.
On the other hand, the necessary luminosity for outgoing jet propagation \citep{2011ApJ...726..107S} is given by
\begin{eqnarray}
L_{iso}\frac{\theta_{op}^2}{2}=
%3.3\times 10^{53}~\mathrm{erg~s^{-1}}
1.3\times 10^{48}~\mathrm{erg~s^{-1}}
\left(\frac{R_0}{10^{11}~\mathrm{cm}}\right)^{4}
\left(\frac{\rho_0}{10^{-5}~\mathrm{g~cm^{-3}}}\right)
\left(\frac{\theta_{op}}{5^\circ}\right)^{2}
\left(\frac{t_{ff}}{1~\mathrm{s}}\right)^{-2}.
\label{eq:L_iso}
\end{eqnarray}
Equating Eqs.  (\ref{eq:L_j}) and (\ref{eq:L_iso}), we get 
\begin{eqnarray}
%2.3\times 10^{-4}=
5.8\times 10^{-3}=
\left(\frac{\eta}{10^{-4}}\right)^{-1}
\left(\frac{t_{ff}}{1~\mathrm{s}}\right)^{-1}
\left(\frac{R_0}{10^{11}~\mathrm{cm}}\right)^{4}
\left(\frac{r}{10^{11}~\mathrm{cm}}\right)^{-3}
\left(\frac{r}{R_0}\right)^{n}
\left(\frac{\theta_{op}}{5^\circ}\right)^{2}.
\label{eq:basic}
\end{eqnarray}
%Note that the normalization of $R_*$ and $r$ is changed.
By deleting $r$ from Eq. (\ref{eq:basic}) using Eq. (\ref{eq:t_ff}) (i.e., $r\sim 10^{11}~\mathrm{cm}(M_r/250M_\odot)^{1/3} (t_{ff}/170~\mathrm{s})^{2/3}$), we can estimate the jet breakout time (i.e., the jet arrival time at $R_0$) as
%Here, we calculate $\lambda$ with varying density index $n$. Our criteria for the successful breakout is that the free-fall time scale should be longer than the breakout timescale, thus
%\begin{equation}
%\frac{t_\mathrm{ff}}{t_\mathrm{b}}\sim
%\end{equation}
%
%We can estimate the jet breakout time with the extended equation from Eq. (12) of \cite{2011ApJ...726..107S} as
%\naga{??? KI thinks the following is correct for the dependence on $M_c$. ???}
%\begin{equation}
\begin{eqnarray}
t_\mathrm{b}(r=R_0)\sim 
%\naga{??6000??}~\mathrm{s}
%6000
%37000~\mathrm{s}
%1.2\times 10^5
170~\mathrm{s}
%\left(\frac{\eta}{10^{-3}}\right)^{-\frac{3}{9-2n}}
%~(8.2)^{-\frac{3}{9-2n}}
\left(\frac{\eta}{10^{-4}}\right)^{-\frac{3}{9-2n}}
\left(\frac{\theta_{op}}{5^\circ}\right)^{\frac{6}{9-2n}}
\left(\frac{R_0}{10^{11}~\mathrm{cm}}\right)^{\frac{3(4-n)}{9-2n}}
%\left(\frac{M_c+0.4M_{\rm env}}{400 M_\odot}\right)^{-\frac{3-n}{9-2n}}.
\left(\frac{M_{R_0}}{250 M_\odot}\right)^{-\frac{3-n}{9-2n}}.
%\mathcal{C}(n)^{-1},
%\nonumber\\
%\times
%\left({10^{-3}}\right)^{-\frac{3}{9-2n}+\frac{3}{3.8}}
%\left({20^\circ}\right)^{\frac{6}{9-2n}-\frac{6}{3.8}}
%\left({10^{13}~\mathrm{cm}}\right)^{\frac{3(4-n)}{9-2n}-\frac{4.2}{3.8}}
%\left({600 GM_\odot}\right)^{-\frac{3-n}{9-2n}+\frac{0.4}{3.8}},
%%%%%%%%%%%
%\left(\frac{M_c}{400 M_\odot}\right)^{-\frac{3}{9-2n}}
%\left(\frac{M_c}{400 M_\odot}\right)^{-\frac{3-n}{9-2n}},
%\end{equation}
%\\
%t_\mathrm{b}\sim 6000~\mathrm{s}
%\left(\frac{\eta}{10^{-3}}\right)^{-1}
%\left(\frac{\theta_{op}}{20^\circ}\right)^{2}
%\left(\frac{R_*}{10^{13}~\mathrm{cm}}\right)
%\end{equation}
\label{eq:tb}
\end{eqnarray}
%where $\mathcal{C}(n)$ is a dimensionless factor as
%\begin{eqnarray}
%\mathcal{C}(n)
%&=&
%%\left({10^{-3}}\right)^{-\frac{3}{9-2n}+\frac{3}{3.8}}
%%\left({20^\circ}\right)^{\frac{6}{9-2n}-\frac{6}{3.8}}
%\left({10^{13}~\mathrm{cm}}\right)^{\frac{3n}{9-2n}-\frac{7.8}{3.8}}
%\left({600 GM_\odot}\right)^{\frac{3-n}{9-2n}-\frac{0.4}{3.8}}
%\left(1~\mathrm{s}\right)^{\frac{3-2n}{9-2n}+\frac{1.2}{3.8}}
%%\left(3\times 10^{-5}~\mathrm{g~cm^{-3}}\right)^{\frac{3}{9-2n}-\frac{3}{3.8}}
%\left(1~\mathrm{g~cm^{-3}}\right)^{\frac{3}{9-2n}-\frac{3}{3.8}}
%\left(\mathrm{g~s^{-1}}\right)^{-\frac{3}{9-2n}+\frac{3}{3.8}}\nonumber\\
%&\approx&
%7.96^{\frac{3-n}{9-2n}-\frac{0.4}{3.8}}
%10^{\frac{11n+84}{9-2n}-\frac{112.6}{3.8}}
%\end{eqnarray}
%which becomes unity for $n=2.6$.
%and the free-fall timescale of matter at the stellar surface with $t_{ff}\approx\sqrt{{R_*}^3/GM_*}$ as
%On the other hand, the free-fall time with Eq. (15) of \cite{2011ApJ...726..107S} as
On the other hand, the free-fall time in Eq.~(\ref{eq:t_ff}) is
\begin{equation}
t_\mathrm{ff}(r=R_0)\sim 
%4000
%1.1\times 10^5~\mathrm{s} 
170
~\mathrm{s}
\left(\frac{R_0}{10^{11}~\mathrm{cm}}\right)^{\frac{3}{2}}
%\left(\frac{M_c+0.4M_{\rm env}}{400 M_\odot}\right)^{-\frac{1}{2}}.
\left(\frac{M_{R_0}}{250 M_\odot}\right)^{-\frac{1}{2}}.
\label{eq:tff}
\end{equation}
%\naga{Note that we have replaced 
%$M_c$ in Eq.~(11) of \cite{2011ApJ...726..107S}
%with  $M_c+0.4M_{\rm env}$ in Eq.~(\ref{eq:tb})
%to match the meaning of the free-fall time 
%in Eqs.~(\ref{eq:tb}) and (\ref{eq:tff}).}
Here, we choose $R_0\approx 0.3 R_*$, beyond which the density profile is not power law but decreases rapidly due to the term of unity in Eq. (\ref{eq:rho}). The jet head accelerates due to decreasing ram pressure of  the ambient matter beyond this radius.
%as well as 
In addition,
the mass accretion rate by the mass shell at $r\gtrsim 0.3 R_*$ decreases rapidly.
Therefore, we consider the critical condition for the successful jet breakout using $R_0\approx 0.3 R_*$ in the following.

By combining Eqs. (\ref{eq:tb}) and (\ref{eq:tff}), 
we obtain the criterion for the successful jet break out, which is that the free-fall time scale should be longer than the breakout timescale, as
%\begin{equation}
\begin{eqnarray}
\frac{t_\mathrm{ff}}{t_\mathrm{b}}\sim 
%0.5
%0.08
1.0
%\left(\frac{\eta}{10^{-3}}\right)^{\frac{3}{9-2n}}
\left(\frac{\eta}{10^{-4}}\right)^{\frac{3}{9-2n}}
\left(\frac{\theta_{op}}{5^\circ}\right)^{-\frac{6}{9-2n}}
\left(\frac{0.3 R_*}{10^{11}~\mathrm{cm}}\right)^{\frac{3}{2(9-2n)}}
%\left(\frac{M_c}{400 M_\odot}\right)^{\frac{3}{9-2n}}
\left(\frac{M_{0.3 R_*}}{250 M_\odot}\right)^{-\frac{3}{2(9-2n)}}
%\mathcal{C}(n)
%\nonumber\\
%\times
%\left({10^{-3}}\right)^{\frac{3}{9-2n}-\frac{3}{3.8}}
%\left({20^\circ}\right)^{-\frac{6}{9-2n}+\frac{6}{3.8}}
%\left({10^{13}~\mathrm{cm}}\right)^{-\frac{3(4-n)}{9-2n}+\frac{4.2}{3.8}}
%\left({600 GM_\odot}\right)^{\frac{3-n}{9-2n}-\frac{0.4}{3.8}}
\gtrsim 1.
\label{eq:tfftb}
%\\
%\frac{t_\mathrm{ff}}{t_\mathrm{b}}\sim 0.5
%\left(\frac{\eta}{10^{-3}}\right)
%\left(\frac{\theta_{op}}{20^\circ}\right)^{-2}
%\left(\frac{R_*}{10^{13}~\mathrm{cm}}\right)^{1/2}
%\left(\frac{M_c+0.4M_c}{600 M_\odot}\right)^{-\frac{1}{2}}
%\gtrsim 1.
%\end{equation}
\end{eqnarray}
According to the notation of Eq. (\ref{eq:tff_tb}), we define $\lambda$ as
\begin{equation}
\lambda\equiv 
%0.5 
%0.030
\left(\frac{0.3R_*}{10^{11}~\mathrm{cm}}\right)^{\frac{3}{2(9-2n)}}
%\left(\frac{M_c}{400 M_\odot}\right)^{\frac{3}{9-2n}}
\left(\frac{M_{0.3R_*}}{250 M_\odot}\right)^{-\frac{3}{2(9-2n)}}.
%\mathcal{C}(n).
\label{eq:lambda}
\end{equation}
As for $M_{0.3R_*}$, we replace this term with $M_c+0.4M_\mathrm{env}$ to estimate the specific values \citep[see][]{2011ApJ...726..107S}.
Using values in Appendix \ref{sec:appendix_a}, 
we find that $\lambda\sim$ 
1.5,
2.6,
2.7,
and 
6.8
%\naga{??? please check these values ???}
for WR, lpop3, mpop3, and RSG, respectively.
\begin{deluxetable}{ccc}
\tabletypesize{\scriptsize}
\rotate
\tablecaption{Summary of stellar models:
(1) Wolf-Rayet star (16TI in \citep{2006ApJ...637..914W}, WR), (2) light Pop III star \citep{2002RvMP...74.1015W} (lpop3), (3) massive Pop III star \citep{2009ApJ...706.1184O} (mpop3).
\label{tab1}} 
\tablewidth{0pt}
\startdata
\hline\hline
 Progenitor model & Total mass ($M_{\sun}$) & Stellar radius (cm) \\
\hline
WR & 14 & $4 \times 10^{10}$  \\
lpop3 & 40 & $1.5 \times 10^{12}$ \\
mpop3 & 915 &$9 \times 10^{12}$ \\
\enddata
%\tablecomments{}
\end{deluxetable}
%%%%%%%%%%%%%%%%%%%%%
\begin{deluxetable}{ccccccccc}
\tabletypesize{\scriptsize}
\rotate
\tablecaption{Summary of our models \label{tab2}} 
\tablewidth{0pt}
\startdata
\hline\hline
 Model & Inner boundary & Enclosed mass &  Efficiency & Half opening angle & Lorentz factor & Specific internal energy & Retarded injection time & AMR level \\
   & $R_{in}$ (cm)& $M_{in} (M_{\sun})$  & $\eta$ & $\theta_{op}$ ($^\circ$)& $\Gamma_{j}$ & $\epsilon_{j}$ & $t_{late}$ (s) &  \\
\hline
WRref & $5 \times 10^{8}$ & 2.0 & $10^{-3}$ & 9 & 400 & $10^{-2}$ & 0 & 3 \\
WRef5-4 & $5 \times 10^{8}$ & 2.0 & $ 5 \times 10^{-4}$ & 9 & 400 & $10^{-2}$ & 0 & 3 \\
WRef2-4 & $5 \times 10^{8}$ & 2.0 & $ 2 \times 10^{-4}$ & 9 & 400 & $10^{-2}$ & 0 & 3 \\
WRef1-4 & $5 \times 10^{8}$ & 2.0 & $ 10^{-4}$ & 9 & 400 & $10^{-2}$ & 0 & 3 \\
WRop3 & $5 \times 10^{8}$ & 2.0 & $10^{-3}$ & 3 & 400 & $10^{-2}$ & 0 & 3 \\
WRop6 & $5 \times 10^{8}$ & 2.0 & $10^{-3}$ & 6 & 400 & $10^{-2}$ & 0 & 3 \\
WRop18 & $5 \times 10^{8}$ & 2.0 & $10^{-3}$ & 18 & 400 & $10^{-2}$ & 0 & 3 \\
WRop36 & $5 \times 10^{8}$ & 2.0 & $10^{-3}$ & 36 & 400 & $10^{-2}$ & 0 & 3 \\
WRop45 & $5 \times 10^{8}$ & 2.0 & $10^{-3}$ & 45 & 400 & $10^{-2}$ & 0 & 3 \\
WRM3 & $5 \times 10^{8}$ & 3.0 & $10^{-3}$ & 9 & 400 & $10^{-2}$ & 7.47 & 3 \\
WRM6 & $5 \times 10^{8}$ & 6.0 & $10^{-3}$ & 9 & 400 & $10^{-2}$ & 26.93 & 3 \\
WRLo5 & $5 \times 10^{8}$ & 2.0 & $10^{-3}$ & 9 & 5 & 60 & 0 & 3 \\
WRLo50 & $5 \times 10^{8}$ & 2.0 & $10^{-3}$ & 9 & 50 & 6 & 0 & 3 \\
WRin25e8  & $2.5 \times 10^{8}$ & 1.7 & $10^{-3}$ & 9 & 400 & $10^{-2}$ & 0 & 3 \\
WRin1e9  & $ 10^{9}$ & 2.5 & $10^{-3}$ & 9 & 400 & $10^{-2}$ & 0 & 3 \\
WRreso5 & $5 \times 10^{8}$ & 2.0 & $10^{-3}$ & 9 & 400 & $10^{-2}$ & 0 & 5 \\
WRreso7 & $5 \times 10^{8}$ & 2.0 & $10^{-3}$ & 9 & 400 & $10^{-2}$ & 0 & 7 \\
WRrepr & $5 \times 10^{8}$ & 2.3 & $10^{-2}$ & 9 & 400 & $10^{-2}$ & 1.94 & 7 \\
WRdelay & $5 \times 10^{8}$ & 2.0 & $10^{-3}$ & 9 & 400 & $10^{-2}$ & 0 & 3 \\
\hline
lpop3ref & $10^{10}$ & 14.9 & $10^{-3}$ & 9 & 400 & $10^{-2}$ & 0 & 3 \\
lpop3ef5e-4 & $10^{10}$ & 14.9 & $ 5 \times 10^{-4}$ & 9 & 400 & $10^{-2}$ & 0 & 3 \\
lpop3ef2e-4 & $10^{10}$ & 14.9 & $ 2 \times 10^{-4}$ & 9 & 400 & $10^{-2}$ & 0 & 3 \\
lpop3ef1e-4 & $10^{10}$ & 14.9 & $ 10^{-4}$ & 9 & 400 & $10^{-2}$ & 0 & 3 \\
lpop3op3 & $10^{10}$ & 14.9 & $10^{-3}$ & 3 & 400 & $10^{-2}$ & 0 & 3 \\
lpop3op6 & $10^{10}$ & 14.9 & $10^{-3}$ & 6 & 400 & $10^{-2}$ & 0 & 3 \\
lpop3op18 & $10^{10}$ & 14.9 & $10^{-3}$ & 18 & 400 & $10^{-2}$ & 0 & 3 \\
lpop3op36 & $10^{10}$ & 14.9 & $10^{-3}$ & 36 & 400 & $10^{-2}$ & 0 & 3 \\
lpop3op45 & $10^{10}$ & 14.9 & $10^{-3}$ & 45 & 400 & $10^{-2}$ & 0 & 3 \\
lpop3in5e9 & $ 5 \times 10^{9}$ & 12.0 & $10^{-3}$ & 9 & 400 & $10^{-2}$ & 0 & 3 \\
lpop3reso7 & $10^{10}$ & 14.9 & $10^{-3}$ & 9 & 400 & $10^{-2}$ & 0 & 7 \\
lpop3repr & $10^{10}$ & 15.1 & $10^{-2}$ & 9 & 400 & $10^{-2}$ & 19.9 & 7 \\
lpop3delay & $10^{10}$ & 14.9 & $10^{-3}$ & 9 & 400 & $10^{-2}$ & 0 & 3 \\
\hline
\tablebreak
\hline\hline
 Model & Inner boundary & Enclosed mass &  Efficiency & Half opening angle & Lorentz factor & Specific internal energy & Retarded injection time & AMR level \\
   & $R_{in}$ (cm)& $M_{in} (M_{\sun})$  & $\eta$ & $\theta_{op}$ ($^\circ$)& $\Gamma_{j}$ & $\epsilon_{j}$ & $t_{late}$ (s) &  \\
\hline
mpop3ref & $10^{11}$ & 414.4 & $10^{-3}$ & 9 & 400 & $10^{-2}$ & 0 & 3 \\
mpop3ef5e-4 & $10^{11}$ & 414.4 & $ 5 \times 10^{-4}$ & 9 & 400 & $10^{-2}$ & 0 & 3 \\
mpop3ef2e-4 & $10^{11}$ & 414.4 & $ 2 \times 10^{-4}$ & 9 & 400 & $10^{-2}$ & 0 & 3 \\
mpop3ef1e-4 & $10^{11}$ & 414.4 & $ 10^{-4}$ & 9 & 400 & $10^{-2}$ & 0 & 3 \\
mpop3op3 & $10^{11}$ & 414.4 & $10^{-3}$ & 3 & 400 & $10^{-2}$ & 0 & 3 \\
mpop3op6 & $10^{11}$ & 414.4 & $10^{-3}$ & 6 & 400 & $10^{-2}$ & 0 & 3 \\
mpop3op18 & $10^{11}$ & 414.4 & $10^{-3}$ & 18 & 400 & $10^{-2}$ & 0 & 3 \\
mpop3op36 & $10^{11}$ & 414.4 & $10^{-3}$ & 36 & 400 & $10^{-2}$ & 0 & 3 \\
mpop3op45 & $10^{11}$ & 414.4 & $10^{-3}$ & 45 & 400 & $10^{-2}$ & 0 & 3 \\
mpop3in5e10 & $ 5 \times 10^{10}$ & 385.1 & $10^{-3}$ & 9 & 400 & $10^{-2}$ & 0 & 3 \\
mpop3reso7 & $10^{11}$ & 414.4 & $10^{-3}$ & 9 & 400 & $10^{-2}$ & 0 & 7 \\
mpop3repr & $10^{11}$ & 431.2 & $10^{-2}$ & 9 & 400 & $10^{-2}$ & 246.6 & 7 \\
mpop3delay & $10^{11}$ & 414.4 & $10^{-3}$ & 9 & 400 & $10^{-2}$ & 0 & 3 \\
\enddata
%\tablecomments{}
\end{deluxetable}
%%%%%%%%%%%%%%%%%%%%%
%%%%%%%%%%%%%%%%%%%%%%
\begin{deluxetable}{ccccccc}
\tabletypesize{\scriptsize}
\rotate
\tablecaption{Summary of our models \label{tab3}} 
\tablewidth{0pt}
\startdata
\hline\hline
 Model & Shock brakout\tablenotemark{a} & Possibility of GRB production \tablenotemark{b}  & $t_{br}$\tablenotemark{c} (s) & $E_{inj}$ ($10^{51}$erg) & $E_{dg}$ ($10^{50}$erg)\\
\hline
WRref & $\bigcirc$ & $\bigcirc$ & 8.20 & 1.67 &  5.60 \\
WRef5-4 & $\bigcirc$ & $\bigcirc$ & 11.78 & 1.23 & 3.79 \\
WRef2-4 & $\bigcirc$ & $\bigcirc$ & 37.09 & 1.71 & 2.36 \\
WRef1-4 & $\times$ & $\times$ & - & - & - \\
WRop3 & $\bigcirc$ & $\bigcirc$ & 6.75 & 1.47 & 5.72 \\
WRop6 & $\bigcirc$ & $\bigcirc$ & 7.22 & 1.54 & 5.65 \\
WRop18 & $\bigcirc$ & $\bigcirc$ & 27.54 & 2.89 & 8.10 \\
WRop36 & $\bigcirc$ & $\times$ & 66.87 & 7.89 & 5.93 \\
WRop45 & $\bigcirc$ & $\times$ & 53.31 & 19.14 & 11.54 \\
WRM3 & $\bigcirc$ & $\bigcirc$ & 7.90 & 1.80 & 6.21 \\
WRM6 & $\bigcirc$ & $\bigcirc$ & 6.18 & 1.54 & 5.60 \\
WRLo5 & $\bigcirc$ & $\bigcirc$ & 8.56 & 1.72 & 5.67 \\
WRLo50 & $\bigcirc$ & $\bigcirc$ & 8.26 &1.68  &5.60  \\
WRin25e8 & $\bigcirc$ & $\bigcirc$ & 6.99 & 1.72 & 5.71 \\
WRin1e9 & $\bigcirc$ & $\bigcirc$ &11.34 &2.05 & 6.17 \\
WRreso5 & $\bigcirc$ & $\bigcirc$ & 7.63 & 1.51 & 4.27 \\
WRreso7 & $\bigcirc$ & $\bigcirc$ & 7.24 & 1.41 & 3.84 \\
WRrepr & $\bigcirc$ & $\bigcirc$ & 3.46 & 6.49 &  27.3\\
WRdelay & $\bigcirc$ & $\bigcirc$ & 8.60 & 1.69 & 5.62 \\
\hline
lpop3ref & $\bigcirc$ & $\bigcirc$ & 288.4 & 2.89 & 11.74\\
lpop3ef5e-4 & $\bigcirc$ & $\bigcirc$ & 454.7 & 1.90 & 7.14 \\
lpop3ef2e-4 & $\bigcirc$ & $\triangle$ & 696.1 & 1.14 & 4.01 \\
lpop3ef1e-4 & $\bigcirc$ & $\triangle$ & 1581.7 & 0.794 & 1.67 \\
lpop3op3 & $\bigcirc$ & $\bigcirc$ & 244.8& 2.90 & 13.5\\
lpop3op6 & $\bigcirc$ & $\bigcirc$ & 257.7& 2.91 & 12.8\\
lpop3op18 & $\bigcirc$ & $\bigcirc$ & 614.1& 2.52 & 8.93\\
lpop3op36 & $\bigcirc$ & $\times$ & 1721 & 2.44 & 6.83\\
lpop3op45 & $\bigcirc$ & $\times$ & 1530.5 & 2.25 & 4.76 \\
lpop3in5e9 & $\bigcirc$ & $\bigcirc$ & 188.9 & 5.63 & 22.57 \\
lpop3reso7 & $\bigcirc$ & $\bigcirc$ &210.5  & 2.38 & 8.72 \\
lpop3repr & $\bigcirc$ & $\bigcirc$ & 118.9 & 11.0 & 47.6\\
lpop3delay & $\bigcirc$ & $\bigcirc$ & 306.6 & 2.94 & 11.49 \\
\hline
\tablebreak
\hline\hline
 Model & Shock breakout\tablenotemark{a} & Possibility of GRB production \tablenotemark{b} & $t_{br}$\tablenotemark{c} (s) & $E_{inj}$ ($10^{51}$erg) & $E_{dg}$ ($10^{50}$erg) \\
\hline
mpop3ref & $\bigcirc$ & $\bigcirc$ & 1219 & 127 & 463 \\
mpop3ef5-4 & $\bigcirc$ & $\bigcirc$ & 1754 & 84 & 269 \\
mpop3ef2-4 & $\bigcirc$ & $\triangle$ & 7213 & 65 & 67 \\
mpop3ef1-4 & $\times$ & $\times$ & - & - & - \\
mpop3op3 & $\bigcirc$ & $\bigcirc$ & 916 & 110 & 514 \\
mpop3op6 & $\bigcirc$ & $\bigcirc$ & 1002 & 116 & 501 \\
mpop3op18 & $\bigcirc$ & $\bigcirc$ & 2620 & 129 & 378 \\
mpop3op36 & $\bigcirc$ & $\times$ & 13427 & 344 & 319  \\
mpop3op45 & $\bigcirc$ & $\times$ & 12263 & 327 & 289 \\
mpop3in5e10 & $\bigcirc$ & $\bigcirc$ & 929 & 147 & 572 \\
mpop3reso7 & $\bigcirc$ & $\bigcirc$ & 1197 & 124  & 410 \\
mpop3repr & $\bigcirc$ & $\bigcirc$ & 563 & 556 & 2437 \\
mpop3delay & $\bigcirc$ & $\bigcirc$ & 1218 & 125 & 475 \\
\enddata
\tablenotetext{a}{\hspace{1mm}$\bigcirc$:Success of shock breakout,\hspace{1mm}$\times$:Failure of shock breakout. }
\tablenotetext{b}{\hspace{1mm}$\bigcirc$:There is a possibility of GRBs,\hspace{1mm}$\triangle$:Touchy to judge the possibility of GRBs,\hspace{1mm}$\times$:There is no possibilities of GRBs.}
\tablenotetext{c}{\hspace{1mm}Time at shock breakout. This time is measured from the beginning of central engine.}
%\tablecomments{}
\end{deluxetable}
%%%%%%%%%%%%%%%%%%%%%
\begin{deluxetable}{cccc}
\tabletypesize{\scriptsize}
\rotate
\tablecaption{The Baryon Mass and Final Lorentz Factor for Jets \label{tab4}} 
\tablewidth{0pt}
\startdata
\hline\hline
 Model & Baryon Mass\tablenotemark{a} ($M_{\sun}$) & Jet Energy\tablenotemark{b} ($E_{j}$) & Average terminal Lorentz factor\tablenotemark{c} ($\Gamma_{f}$) \\
\hline
WRref & $9.25 \times 10^{-5}$ & $9.43 \times 10^{51}$ & 57 \\
WRreso5 & $5.36 \times 10^{-5}$ & $9.70 \times 10^{51}$ & 101\\
WRreso7 & $2.61 \times 10^{-5}$ & $9.78 \times 10^{51}$ & 209\\
WRM3 & $7.43 \times 10^{-5}$ & $8.42 \times 10^{51}$ &  63 \\
WRM6 & $4.57 \times 10^{-5}$ & $5.73 \times 10^{51}$ &  70\\
WRrepr & $2.52 \times 10^{-5}$ & $9.17 \times 10^{52}$& 2035\\
%\hline
%lpop3ref & $1.61 \times 10^{-4}$ \\
%lpop3reso7 & $5.01 \times 10^{-5}$ \\
%lpop3repr & $7.17 \times 10^{-5}$ \\
%\hline
%mpop3ref & $3.62 \times 10^{-3}$ \\
%mpop3reso7 & $9.71 \times 10^{-4}$ \\
%mpop3repr & $7.82 \times 10^{-4}$ \\
\enddata
\tablenotetext{a}{\hspace{1mm} The baryon mass contained in the jet at the breakout. See Eq.~(\ref{eq:baryonmass}) in Section~\ref{subseclimitation} for its definition.}
\tablenotetext{b}{\hspace{1mm} The total energy of injected jet after the breakout.}
\tablenotetext{c}{\hspace{1mm} The average terminal Lorentz factor of outflows estimated by Eq.~(\ref{eq:averagetermgamma}).}
%\tablecomments{}
\end{deluxetable}

%%%%%%%%%%%%%%%%%%%%%%%%%%%%%%%%%%%%%%%%%%%%%%%%%%%%%%%
%%%%%%%%%%%%%%%%%%%%%%%%%%%%%%%%%%%%%%%%%%%%%%%%%%%%%%%
%%%% add revise 1
%%%%%%%%%%%%%%%%%%%%%%%%%%%%%%%%%%%%%%%%%%%%%%%%%%%%%%%%%
% dependence of efficiency
%%%%%%%%%%%%%%%%%%%%%%%%%%%%%%%%%%%%%%%%%%%%%%%%%%%%%%%%%%%%

% dependence of opening angle

% Injection timiing

\clearpage

%% \begin{figure}
%% \vspace{15mm}
%% \epsscale{1.0}
%% \plotone{ftempo4.eps}
%% \caption{Time evolutions of Lorentz factor on z-axis for each representative model. The reverse shock is located where the Lorentz factor suddenly drops from 400 to several. 
%% \label{ftempo4}}
%% \end{figure}

% inner boundary and resolution dependence

%%%%%%%%%%%%%%%%%%%%%%%%%%%%%%%%%%%%%%%%%%%%%%%%%%%%%%%

%\begin{figure}[htbp]
%%\vspace{15mm}
%\epsscale{0.8}
%%\plotone{M-R.eps}
%\caption{The schematic picture of the possible GRB progenitor in the stellar mass-radius diagram \naga{by the analytical estimate}. The white region is feasible progenitor. The circles represent the position of each progenitors considered in this paper. The bottom grey region is unfavorable because the time necessary for jet propagation toward the stellar surface is longer than the active timescale of the central engine (i.e., the mass accretion timescale), on the other hand the top grey region is also unfavorable because the velocity of the jet head is slower than that of almost spherical cocoon component.  We obtain these lines with $\eta=10^{-3}$, $\theta_{op}=20^\circ$, $n=2.6$, and $M_c=M_\mathrm{env}=0.5M_*$.
%As for the WR star, our assumptions of simple model in \cite{2011ApJ...726..107S} are invalid.
%\label{f21}} 
%\end{figure}

%%%%%%%%%%%%%%%%%%%%%%%%%%%%%%%%%%%%%%%%%%%%%%%%%%%%%%%

%% \begin{figure}
%% \vspace{25mm}
%% \hspace{-25mm}
%% \epsscale{0.7}
%% \plotone{f2.eps}
%% \caption{
%% The radial profiles of the electron temperature $T_e$ [K] for one snapshot ({\it dots}) and the time-averaged data ({\it thick dashed line}) for our simulation data. The data for the snapshot is same as those used in the third column of Figure~\ref{f1}. We also show the electron temperature profile for the RIAF model used in \cite{bl09} ({\it dotted line}). 
%% \label{f2}}
%% \end{figure}

\end{document}